# Only-child matching penalty in the marriage market[1]

Keisuke Kawata[2]

Mizuki Komura[3]

July 25, 2023

This study explores the marriage matching of only-child individuals and its outcome. Specifically, we analyze two aspects. First, we investigate how marital status (i.e., marriage with an only child, that with a non-only child and remaining single) differs between only children and non-only children. This analysis allows us to know whether people choose mates in a positive or a negative assortative manner regarding only-child status, and to predict whether only-child individuals benefit from marriage matching premiums or are subject to penalties regarding partner attractiveness. Second, we measure the premium/penalty by the size of the gap in partner's socio economic status (SES, here, years of schooling) between only-child and non–only-child individuals. The conventional economic theory and the observed marriage patterns of positive assortative mating on only-child status predict that only-child individuals are subject to a matching penalty in the marriage market, especially when their partner is also an only child. Furthermore, our estimation confirms that among especially women marrying an only-child husband, only children are penalized in terms of 0.57-years-lower educational attainment on the part of the partner.

[1] We are grateful to Alessandro Cigno, Chishio Furukawa, Taiyo Fukai, Haruaki Hirota, Hiroyuki Kasahara, Nobuyoshi Kikuchi, Shinnosuke Kikuchi, Ayako Kondo, Hisaki Kono, Miki Kohara, Annalisa Luporini, Yuki Masujima, Takahiro Miura, Kazutoshi Miyazawa, Takeshi Murooka, Akira Nishimori, Hikaru Ogawa, Naohiro Ogawa, Shinpei Sano, Takashi Shimizu, Kenta Tanaka, Kensuke Teshima, Takashi Unayama, David Weil and Junichi Yamasaki for their constructive and insightful comments. We would like to thank the participants of Workshop on the Economics of Low Fertility and Aging, Tokyo Labor Economics Workshop, the Kansai Seminar for Studies on Labor, Kyoto Summer Workshop on Applied Economics, and seminars at Kyoto University, The Univesrsity Tokyo, Kobe University (Rokko Forum). This study is supported by JSPS KAKENHI (Grant Numbers 18K01661, 22H00854 and 23K01418). This study was conducted based on data provided by the National Institute of Population and Social Security Research (NIPSSR) under Article 33 of the Statistics Law. There are no conflicts of interest to declare. All errors are our own.

[2] Institute of Social Sciences, University of Tokyo, 7-3-1 Hongo, Bunkyo-ku, Tokyo 113-0033, Japan, Email:keisukekawata@iss.u-tokyo.ac.jp

[3] School of Economics, Kwansei Gakuin University, 1-155, Uegahara Ichibancho, Nishinomiya 662-8501, Japan, Email:m.komura@kwansei.ac.jp

## 1. Introduction

Becker's marriage matching theory can describe who marries whom or remains single and the amount of marital gain (Becker 1973, 1974). Marriage patterns with respect to the marital partners' characteristics affect not only the subsequent welfare of the individual but also income inequality between and within households. They are also linked to intergenerational mobility, child reproduction, and economic growth and may even be related to discrimination issues. Recently, there has been accumulating evidence on marriage matching, focusing on the specific factors that can affect the attractiveness of each candidate. Their results show that marriage matchings are influenced by acquired characteristics and even innate traits such as race or sexual orientation.[4] One characteristic that is outside one's control is sibling composition. Exceptionally, only-child individuals have unique features, irrespective of their cultures or certain social norms.

Whether one can marry and to whom is hardly a matter of concern only for the only child: it also significantly impacts intergenerational relations. One of the major inducements for people to marry is economies of scale (Browning, Chiappori, and Weiss (2014)).[5] Many assume that they will be single or two in most cases, but if we take a larger view of the family, whether the marriage partner is strictly an only child will significantly impact the size of each natal family. While the only child is young, the parents can devote many resources to this dependent child. However, when parents become old and dependent, they cannot benefit from economies of scale. Unlike those with siblings, only children face the task of caring for their parents alone, typically after their prime marrying age.[6] While the labor market may work and resolve these intergenerational burden gaps among regions or nations, the marriage market may act on those among families. How does the marriage market affect this disparity between only-child and non-only-child households?

---

[4] The studies have examined race (Pierre-André Chiappori, Oreffice, and Quintana-Domeque 2016), sexual orientation (Jepsen and Jepsen 2002), obesity (Pierre-André Chiappori, Oreffice, and Quintana-Domeque 2012), education (Choo 2015; Siow 2015), and smoking habits (Pierre-André Chiappori, Oreffice, and Quintana-Domeque 2012).

[5] Strictly, it is said that having a larger family comes with numerous benefits, such as the provision of public goods, risk sharing, and the advantages of economies of scale.

[6] Indeed, the literature has proved that strong family ties or customs might negatively impact the younger generation's economic activities, especially in industrialized and urbanized economies (Alesina and Giuliano 2010). In families with only children, where externalities cannot occur, children are less likely to leave their parents (Konrad et al. 2002a; Rainer and Siedler 2009), resulting in fewer opportunities in the labor market (Rainer and Siedler 2009).

We have seen a global increase in only-child families in many developed countries. For example, in the American and Asian spheres, the percentage of one-child families among those with children has nearly doubled in recent decades.[7] According to Eurostat (2022), the percentage in European countries is almost half at 49% in 2021, while The Office for National Statistics (2020) reports a figure of 43.7% for the UK in 2019. These social trends are affected by many modern issues, such as economic concerns, higher age of becoming a parent, infertility, marriage lives and careers with high pressure, the growing expense of raising children, and the simple desire to have only one child. China's one-child policy (OCP) has also contributed to the number of only-child families worldwide. There is, however, little understanding of only-child individuals' marriage outcomes.

The purpose of this paper is twofold: First, we investigate the marriage patterns for only-child individuals. In the study, we compare patterns in the likelihood of marriage of only-child individuals, excluding the scale effects between only-child and non–only-child individuals (Pierre-André Chiappori, Oreffice, and Quintana-Domeque 2012). Specifically, we examine how marital status (i.e., marriage with an only child, that with a non-only child, and remaining single) differs between only children and non-only children, controlling for age, sex, and birthplace. This comparison allows us to understand only children's assortativity and predict whether an only-child matching premium or penalty exists in the marriage market. Second, we measure the predicted only-child marriage matching outcomes following Pierre-André Chiappori, Oreffice, and Quintana-Domeque (2018), where the matching premium/penalty is measured by the difference in the partner's attractiveness.

We make several contributions to the literature by investigating marriage matching in Japan. On the one hand, this study explores the role of the marriage market by clarifying the nature of assortativity on sibling composition. If the burden of caring for parents differs between only children and non-only children, the message that the degree of assortativity sends in the context of an aging society is significant. For example, suppose that the marriage market is characterized by negative assortative mating (i.e., only and non-only children are more likely to marry each other). In this case, the burden of family caregiving is moving toward equalization. In contrast, in the case of positive assortative mating, the marriage market is accelerating inequality in this aspect. Therefore, our answer may be meaningful for ascertaining whether the marriage market is driving equalization in family size. The latest sociological studies have started to emphasize the significance of a sibling position in marriage with strong parental caretaking responsibilities and observes the lower likelihood of marriage with only children, first sons, and first daughter without male siblings (Yu and Hertog (2018); Uchikoshi, Raymo, and Yoda (2023)). We contribute to the social sciences from an economic perspective to clarify the role of the marriage market by looking at the nature of one-child marriages as well as their outcomes in the market as a whole.

---

[7] For instance, the percentage in the US increased from 11% in 1976 to 21% in 2016; from 12% of families in Canada in 1981 to 26% in 2019; and from 10% in 2002 in Japan to 18.6% in 2015. In Singapore, 19.0% of married women had one child in 2010, but 24% did so in 2020.

In addition, this is the first study to measure marriage quality for only-child individuals. As noted earlier, there are few studies on marriage match quality with respect to only-child status or sibling composition, which are innate characteristics. Yu, Su, and Chiu (2012) and Angrist, Lavy, and Schlosser (2010) look at the effects on the marital status and age at first marriage of only children as the marriage outcomes. Despite their insightful findings, these works fail to capture the perspective of the marriage market and matching with a partner. Thus, it is difficult to discern whether the individual chooses his or her marital status or is forced to stay single. Vogl (2013) examines women's arranged marriages in Nepal and the impact of sibling presence on marriage outcomes and partner quality. Our study complements the literature by shedding light on only-child marriages in developed countries with low fertility and aging populations from the marriage market candidates' perspective. While the marriage partner's sibling composition and educational background each provide information that characterizes the marriage of an only child, we can link the two results with a standard decision model of marriage matching. If we assume a two-dimensional decision-making model that evaluates attractiveness in the marriage market in line with recent studies on marriage matching (e.g., Pierre-André Chiappori, Oreffice, and Quintana-Domeque 2012, 2018), we can obtain the implication for the situation that only children face in in the marriage market.

Exploring the marriage matching of only children in Japan represents more than an exercise of academic curiosity. Many developed countries are experiencing population aging, with declining birth rates accompanied by an increasing number of only children. These phenomena have become a social problem as the burden of caring for elderly individuals, with their longer life spans and extended caregiving periods, falls on their fewer children. Rainer and Siedler (2012) also suggest that the burden on the only child depends on the strength of social security and social expectations for informal family care. Obviously, Japan is one of the countries with the most severely aged population and has an established social security system. At the same time, however, the norm of filial obligation toward one's parents tends to be strong in Japan, partly due to the spread of Confucianism in Asia (which will be discussed later in this paper). Therefore, if we can highlight only children's relatively solid bond with their parents as a possible mechanism behind only-child marriage matching outcomes,

Japan is an interesting arena to bring the possible mechanisms of the interdependence of parent-child relationships and marriage to light. Only children in the Japanese marriage market are also an appealing research population from the perspective of external validity. A unique policy, OCP, rapidly increased the number of only children in China, and many implications can have been drawn from this exogenous shock for each household. However, when there is a uniform increase in the number of only children in the same generation, it is not ideal to analyze the choice of either an only child or a non-only child as a marital partner based on the marriage matching theory. Thus, we need to target a society where only children and children with siblings coexist in the marriage market in the same cohort.

According to our first exercise on assortativity in marriage patterns of pairing, only children have different patterns from non-only children. First, only children are less likely to be married. Second, the likelihood of marrying an only-child partner is higher for only children but lower for marriages with a non-only-child partner, corresponding to positive assortativity on only-child status. Then, the canonical model and the observed marriage patterns of pairing predict an only-child marriage matching penalty, where only-child individuals give up on partner attractiveness, as determined by characteristics other than only-child status, given their lower outside option in the marriage market. We further expect on the basis of our finding of positive assortative mating in only-child status that this effect is more prominent for marriages between only children. Finally, testing our hypothesis of an only-child matching penalty, we confirm that it exists in the marriage market and that the magnitudes of the penalty seem gender asymmetric and depends on the partner's only-child status, where the latter is consistent with our theory and the findings on marriage patterns of pairing. Specifically, while the only-child matching penalty was not observed in the pooled sample, only-child women suffer the severest matching penalty in the form of a reduction in their partner's education of approximately 0.57 when they marry a man who is also an only child; this magnitude is comparable to or even larger than the size of the gender gap in education in our sample at 0.54.

In addition, we conduct two further analyses: First, we analyze the effects of heterogeneities in their own birth year, age, and own education. Second, to consider the effects of alternative sibling positions relative to that of only-child families, we measure the matching penalty for heirs characterized by two definitions of patrilineal and primogeniture. We find stronger assortativity in only-child status among the marriages in recent years, but that own higher education can mitigate the assortativity. However, these heterogeneities do not affect the main results of the partner's academic background. On the other hand, the penalty effects for heirs in both definitions are also consistently observed to be smaller than the only-child matching penalty effect.

The rest of the paper is organized as follows. First, Section 2 reviews the related literature, while Section 3 explains Japan's relevant background. Then, Section 4 shows the details of the data that we use. Section 5 presents the conceptual framework that our study relies on. Section 6 elaborates on the marriage patterns based on only-child status and provides hypotheses for matching outcomes. Section 7 measures the marriage-matching outcomes based on the hypotheses. In Section 8, we further conduct supplementary analyses, and in Section 9, we conclude the paper.

## 2. Related literature

This study contributes to a better understanding of the marriage patterns of only children and, by extension, of lifetime welfare. As social scientists from various disciplines seek to better understand the effects of sibling composition (birth order), our research relates to a wide range of relevant disciplines and literature. This section provides an overview of previous studies examining the relationship between sibling structure and marriage. Extensive research has been conducted on the correlation between sibling structure and an individual's educational background, as well as on the relationship between sibling structure and various other characteristics. The appendix A delves into a discussion of these latter two aspects.

While literature has accumulated on the impact of sibling structure on individual welfare, focusing more on its psychological and socioeconomic aspects, its impact on marriage has remained relatively unexamined. In regions with cultural norms that strongly bind adult children and their parents, such as East Asia and Southern Europe (Cordón 1997; Giuliano 2007; Raymo and Ono 2007; Yu and Kuo 2016), sibling composition may be particularly relevant to marriage outcomes including partner choice. Since little is known about the association between sibling structure, particularly only-child status, and family formation, studying this relationship can enhance our understanding of both lines of literature.

There are only a handful of exceptions that look at marriage outcomes directly. In the sociology literature, Yu, Su, and Chiu (2012) explore the effects of sibship size, birth-order rank, and sibship gender composition on one's age at first marriage. They find asymmetric results by gender, which they interpret to be attributable to the social norm of gender roles in families. Specifically, men with no male siblings marry earlier, while women marry earlier when they have more siblings or are in earlier birth positions. In a Japanese study, Kojima (1993) examines the marital arrangement as a marriage outcome based on various hypotheses based on sibling composition. Economists have also started to focus on marriage outcomes. For example, Angrist, Lavy, and Schlosser (2010) consider the marriage outcome of quality of children, showing that those with many younger siblings tend to be married, to marry earlier and to have more children (though the sample does not include only children).

Recent sociology has begun to point out the significance of the intergenerational relationship of individuals in specific sibling composition on romantic and marriage outcomes, sharing similar motivations as ours. Yu and Hertog (2018) analyzes the effect of sibling composition, including only children, on the online dating market, assuming that being the eldest or only child in a Confucian society like Japan signals their caregiving obligation to the potential marriage partner. The analysis found that only children are less likely to find a partner. Using representative data of Japan, Uchikoshi, Raymo, and Yoda (2023) also analyzes the effect of the change in population structure of sibling composition (including only children) on demographic change, using different indicators to show that children who are expected to care for their parents are less likely to marry. Specifically, they calculate the percentage of the male/female population with a particular sibling composition that is actually married to that combination.

Previous studies have also considered the relationship between parental relationships and family formation. A strong parent–child relationship will have an impact early and late in adolescence. For example, many sociologists have found that children who live with their parents tend not to marry or to marry at a higher age (Raymo 2003; Raymo and Ono 2007; Sakamoto and Kitamura 2007; Yu and Kuo 2016). They interpret this result as indicating that a strong relationship with parents can make children reluctant to marry because of economic incentives (i.e., the income effect), the availability of a comfortable home with support, or psychological readiness for marriage. Thus, it is difficult to give an interpretation to the difference in the marital status or timing of such children's marriages In other words, we cannot understand whether these individuals remain in the marriage market of their own accord or are forced to make such choices under unfavorable conditions.

Several solutions to this problem exist. The first is to look directly at the process of couple formation. The Yu and Hertog (2018)'s study of online dating of only children may help interpret their subsequent marriages. They conclude that only children are disadvantaged because only children are more likely to be senders than receivers of requests among registrants willing to date and that that their requests are less likely to be approved. We complement Yu and Hertog (2018) by comparing the actual marital partner's quality and sibling composition between only and non-only children as a second solution. Vogl (2013) is the only study that looks at marriage quality by focusing on sibling composition in the veiw of economics. He looks at women's arranged marriages in Nepal, a developing country with a growing population. He shows that women are rushed into marriages due to the presence of younger sisters, resulting in marriages with less-qualified partners. In addition to looking at partner's quality, we theoretically induce the implication for the latent on the condition of marriage market faced by only children using the two observed matching outcomes of assortativity and the partner's SES.

Moreover, the inequality that we can address is not limited to SES: we can also consider the burden of caregiving. Recent studies have raised theoretical questions about whom individuals from families with strong norms of filial obligation tend to marry. Unlike other transmitted values factors such as ethnicity, religion, attitudes towards working women, or economic preferences (Bisin and Verdier (2000); Bisin, Topa, and Verdier (2004); Fernández, Fogli, and Olivetti (2004);Wu and Zhang (2021)), the transmission of strong family norms regarding filial piety can potentially create conflicts within families. According to Cigno, Komura, and Luporini (2017) and Cigno, Gioffré, and Luporini (2021), individuals from families with strong filial piety norms may choose to marry partners who also uphold these norms in order to preserve them. On the other hand, it is also possible for them to prioritize their own family by marrying partners with weaker norms. If we consider only children as the group with the strongest parental care norms, assessing assortativity can help us provide some insights into this question. Although we do not directly uncover the underlying mechanism behind these results, our findings on partner choices can offer a certain level of understanding through observable outcomes in the younger generations' adult pairings.

Finally, we contribute to the marriage matching literature. Becker's application of matching theory to marriage enabled marriage market analysis. The discovery of positive assortativity on education and SES also revealed the impact on household inequality (e.g., Mare (1991); Pencavel (1998); Fernández and Rogerson (2001); Breen and Salazar (2011); Breen and Salazar (2011); Greenwood et al. (2014); Greenwood et al. (2016); Eika, Mogstad, and Zafar (2019)). In addition, by demonstrating matching patterns, studies have provided some evidence of discrimination at marriage on certain traits that the individual cannot change. The form of sibling composition addressed in this study is also not determined by the individual's own will. There is a trend toward extending the measure of the attractiveness of marriage market participants in empirical analyses from including just one dimension to covering multiple dimensions (Pierre-André Chiappori, Oreffice, and Quintana-Domeque 2012, 2018). Pierre-André Chiappori, Oreffice, and Quintana-Domeque (2018) utilize personal traits and SES to measure matching penalties and premiums. This study contributes to the field by using sibling structure as an application and successfully measures the matching outcome (premium or penalty) based on this theory after clarifying the matching patterns in the marriage market.

# 3. Background

This chapter provides background on the distinctive character of the Japanese family and the only child.

## 3.1 Families in Japan

Confucianism has deeply affected the family system in Asian countries, where repaying parents is considered a virtue, as the philosophical rationale for strong family ties is traced to Catholicism in Europe (Esping-Andersen (1997)), resulting in a relatively large reliance on families than society. In Japan, the duty of filial piety remains relatively strong, and caring for elderly parents has traditionally been a family affair. Japanese law lists parents' financial support as an obligation of adult children. Article 877(1) of the Civil Code stipulates that "immediate blood relatives and brothers and sisters are obligated to support each other," and parents are included among these immediate blood relatives. However, in principle, this support refers to transferable financial support, and the support obligation does not mean that one must live with and care for one's parents.

In reality, however, many people recognize that caring for parents is also an obligation of children. Ogawa and Retherford (1993) see changes in both norms of filial obligations and expectations for children to care for themselves. Expectations toward children have declined sharply over the years (with 18% of parents reporting that they "expect" support from their children, 20% that they "never thought about it," and 62% that they "do not expect" support in 1990). Conversely, although the authors conclude that attitudes toward norms have weakened since 1986, 72% of respondents accept the norms, characterizing filial support as a "good custom," "natural duty," or "unavoidable" in 1990, the most recent year for which results are available (12% answered "bad norm," and 15% answered "other").

(**population2020?**) reported that as of 2018, 19.8% of households whose wife is under 70 live with one of their four parents. While the figure shows a decrease from the 1998 survey result of 26.2%, it is still higher than that in other major industrialized countries (6.2% in the United States, 3.4% in Germany, and 1.5% in Sweden in 2015).[8] Moreover, according to a report by the Ministry of Health, Labour and Welfare (2020), 28.2% of the primary caregivers for elderly people requiring long-term care (LTC) were coresident couples of the younger generation (the elderly people's children + children's partners) in 2019.[9] This figure is more than double the 12.1% of the care provided formally (by paid caregivers), indicating that the burden of care on the younger generation is still not being shouldered by the market.

### 3.2 Patrilineal Lineage and Heirs in the Japanese Families

Until the end of WWII, the inheritance system in Japan was patrilineal, with the eldest son inheriting the entire estate as the family head. In 1947, the law was amended to allow family members other than the heir (typically the eldest son) to inherit equally. However, for families that have existed for a long time or farm families, there is still a culture of inheritance by and associated heavier obligations on the family head (i.e., eldest son).

[8] The Japanese Cabinet Office conducted an international comparative survey in 2015 covering Japan, the U.S., Germany, and Sweden and asked 2,800 men and women aged 60 and older (excluding institutionalized residents) about various issues in the lives of elderly individuals. In 2015, 14.8% of respondents to this survey in Japan said that they have children living with them.

[9] Comprehensive survey of living conditions 2019.

Moreover, the burden of care tends to be on a specific child and their spouse, or the children's couple living with their parents. If the child living with his parents is the eldest son of the heir, the woman who will become his wife will make her marriage decision based on the assumption that she will take care of her parents-in-law. Ogawa and Ermisch (1996) show that this solid family structure in Japan strongly influences women's employment in the younger generation who live together with their parents (or, in the case of male children living with parents, their in-laws). Although they are more likely to work full-time when younger, they are less likely to be paid workers if they have to care for a parent or parent-in-law in or near the household. As a result, the authors find that time spent caring for an elderly relative decreases women's wages in their full-time jobs. It also decreases the probability that they will find full-time paid employment while increasing their probability of finding part-time work or becoming housewives.

In addition to provision for elderly care, heirs were supposed to pass down blood, surnames, and business and maintain ancestral graves. According to Wakabayashi and Horioka (2009), family trends were generally consistent with the dynasty hypothesis, suggesting a strong relationship between the heir and the parents. Self-employed parents were more likely to live with their children, and those who took the wife's name were less likely to live with the husband's parents. Even if the eldest child is female, she tends to live with the eldest son, suggesting that the cultural centrality of the eldest son is still strong. When the eldest son is considered as a marriage partner, it is necessary to understand the associated familial role of the eldest couple.

However, it should be mentioned that the masculine norm is slowly fading and that the tendency to care for one's own parents has increased. According to (**population2020?**), as of 2018, the strong tendency under patrilineal culture to live with the husband's parents continues today. The organization reports the percentage of households in which the wife is under 70 living with either the wife's or the husband's parents, with 3.8% of the respondents living with the wife's father, 5.7% with the wife's mother, 10.5% with the husband's father, and 13.2% with the husband's mother. A comparison with the figures of 6.5% living with the wife's parents and 22.2% living with the husband's parents in 1998 suggests that the gap in the percentage living with the husband's parents and the wife's parents is shrinking. Thus, the norm of giving priority to parents on the husband's side is fading, although it still exists in the society.

### 3.3 Only children in Japan

A survey conducted since 1940 shows that the percentage of only-child families has gradually increased since the 1990s (Cabinet Office 2021). The ratio of only-child families among those with children has increased from 10% in 2002 to 18.6% in 2015. Other trends show that the percentage of mothers having two children has remained unchanged at more than 50% throughout almost 40 years. The increase in the share of households with only children can also be attributed to the decrease in households with three or more siblings since the early 2000s.

Given Japan's male-dominated society, where strong family norms remain, and the burden is concentrated on the offspring, especially a specific child, an only male or female child is a unique characteristic in post-marital life (Yu and Hertog (2018); Uchikoshi, Raymo, and Yoda (2023)).[10] When an only child reaches adulthood, he or she automatically become the parents' heir. If the only child is a male, he automatically becomes the eldest son. If the only child was female, there would be no eldest son to serve as the typical successor and no other siblings to share the burden in their natal family. Historically, in the case of female-only-child families, as there were no males among the children, a system of adoption of the son-in-law was commonly used to keep the blood or surnames (Wakabayashi and Horioka (2009)). Otherwise, an only-child woman would marry into her husband's family and prioritize the care of her in-laws over her own parents. As a result, the natal families of only-child women had to give up at their marriage the intergenerational relation as well as their surnames and family lines which had been preserved over generations. In the trend toward masculine domination is weakening these days (as discussed earlier), however, only children will bear the burden of caring for their own parents irrespective of gender, especially in the younger generation.

It should also be noted that an only child has the advantage that they can enjoy the transfers from their parents throughout their lives. One prime example of post-marital income transfers is bequests from their parents. According to The Yu-cho Foundation (2023), the percentage of respondents who divided their inheritance equally among siblings when they received it from their parents has remained stable at around 50% for about ten years since 2013. Moreover, approximately 60% of the respondents said they would divide their estates equally among their children according to the number of children. In addition, as approximately 10% said they would leave all of their estates to their only child, the estate received by the only child is expected to be larger. According to The Dai-ichi Life Research Institute (2007), with 715 samples, the average inheritance from parents is 21.86 million yen, 14.05 million yen for the eldest child, and 13.03 million yen for the second and subsequent children. Combined with the discussion of investment in education, which will be discussed later, only children seem not only disadvantaged but benefit from intergenerational relations.

---

[10] The detailed arguments are also found in Yu and Hertog (2018) and Uchikoshi, Raymo, and Yoda (2023).

## 4. Data and summary statistics

### 4.1 The National Survey on Migration

The National Survey on Migration is a national representative survey in Japan. The data contain information on sibling configuration, birthplace prefecture, year of birth, and marital status of each family member. The data is collected by the National Institute of Population and Social Security Research through prefectural stratified sampling from the survey areas established in the National Survey of Living Standards. The surveyors distribute and collect the questionnaires for each household. This study uses the latest available waves from 1991, 1996, 2001, 2006, and 2011.[11] The collection rates for each wave are 89.4%, 95.8%, 85.5%, 74.0%, and 74.7%, respectively. Unfortunately, the weight is not provided for researchers, so our data is left unadjusted.

The sample of singles is restricted to persons who have never been married. We restrict the sample of married couples to married persons, excluding divorced, bereaved, and separated persons. Furthermore, we use the information on the household heads and their spouses aged 23 to 65 in the responding year for whom complete educational background information is available. The latter threshold is due to the definition of the variable for sibling structure. Siblings in this questionnaire are limited to those still living. By setting the maximum age for the sample, we take into account the possibility that older survey respondents may have lost a sibling after marriage due to life expectancy. In addition, we restrict the sample to married couples with information on sibling configuration, year of birth, and hometown prefecture. The final sample includes 46,981 observations.

**Only child dummy:** The survey asks for the number of surviving older brothers, older sisters, younger brothers, and younger sisters. Those with none of these present are assigned a value of 1 for the only child dummy. All others are assigned a value of 0.

---

[11] The sample size for the 1991 wave relatively is small. This is mainly due to missing values for the place of origin. In addition, the 1991 wave uses slightly different questionnaires, although it has our necessary variables, whereas the other waves have been changed to be more uniform.

**Age at the time of the survey:** To account for changing trends over five-year intervals, we controlled for the age of respondents at the time of the survey along with the information on the year of birth.

**Years of schooling:** The number of years of education for the individual is calculated from the highest school category.

**Birth year:** The person's year of birth is used.

**Regional block:** The survey asks for the prefecture where the respondent was born. In the analyses, 47 prefectures are classified into ten blocks and used to take into account regional characteristics.[12]

## 4.2 Summary statistics

Table 1 is around here

Table 1 presents the descriptive statistics by respondents' gender and only-child status. The numbers indicate the medians within each subsample. The brackets show the percentage or 25th and 75th percentiles within the subsample. What we can see from this is as follows. First, the sample size of only-child individuals is necessarily small. Second, the marital status of only children is more likely to be single than non-only children at the descriptive level. Third, when we look at education, our correlate of interest, while the values are the same at the mean level, there is a difference depending on only-child status: only children tend to be slightly more educated. Finally, the mean value of only children's birth year is larger than non-only children. This may reflect the recent trend of increasing only-child families.

Table 2 is around here

Next, let us look at partners' characteristics. Table 2 demonstrates descriptive statistics of spouses for married individuals by respondents' gender and only-child status. Again, as scale effects are at work, the probability that the partner is an only child is small. However, when the respondent is an only child, the likelihood of marrying an only-child spouse is approximately three times greater than that of marrying a non–only-child spouse in a simple comparison. The marital partners' years of schooling remain similar at the mean level as in the descriptive statistics but again slightly higher for only children.

---

[12] Specifically, Hokkaido for the "Hokkaido" block; Aomori, Iwate, Miyagi, Akita, Yamagata, and Fukushima for the "Tohoku" Block; Saitama, Chiba, Tokyo and Kanagawa for the "Minamikanto" block; Ibaraki, Tochigi, Gunma, Yamanashi, and Nagano for the "Kitakanto and Koshin" block; Niigata, Toyama, Ishikawa, and Fukui for the "Hokuriku" Block; Gifu, Shizuoka Aichi, and Mie for the "Tokai" block; Shiga, Kyoto, Osaka, Hyogo, Nara, Osaka and Wakayama for the "Kinki" block; Tottori, Shimane, Okayama, Hiroshima and Yamaguchi for the "Chugoku" Block; Tokushima, Kagawa, Ehime, and Kochi for the "Shikoku" block; Fukuoka, Saga, Nagasaki, Kumamoto, Oita, Miyazaki, Kagoshima, and Okinawa for the "Kyushu and Okinawa" block.

Before we move on to further analysis, let us discuss the key variable, years of schooling (i.e., a proxy of SES in our study). The descriptive statistics reveal that only children in our sample tend to have a relatively higher level of education. Some may argue that using one's own education as a control variable can pose challenges to making accurate causal inferences.[13] However, the objective of this study is not to measure the causal effects but rather to shed light on how the presence or absence of siblings influences family formation among adults. Therefore, this study includes one's own education as a control variable when investigating differences in marriage patterns. There are several reasons for this: firstly, we acknowledge that there is positive assortativity in partner selection based on educational backgrounds, which is one of our main interests in this analysis. Secondly, previous studies have also taken into account one's own education as a control variable (Pierre-André Chiappori, Oreffice, and Quintana-Domeque (2018)). Furthermore, even if education is not controlled for, it does not significantly alter the main findings based on our economic model (see Section 8). Hence, we consider the controlled analysis as the primary outcome of this study.

---

[13] Considering the reasons for having an only child with a focus on parental characteristics that may influence the children's SES, being an only child may have both a positive and negative relationship with an individual's SES. A negative relationship could simply be an income effect, where the parents are not wealthy enough to have a second child. The divorce of the parent couple while the child is young may also forestall the birth of a second child. As a positive relationship, the number of children may be reduced to increase the resources per person, reflecting the quality mechanism (in the quantity–quality trade-off) instead of the income effect. It is also possible that the age of first childbirth has increased because mothers have accumulated higher educational investments of their own. Fairly strong assortativity on education has been observed between spouses. In addition, parents may reduce their number of children due to a lack of space in urban areas, as a regional-specific factor other than the parents' SES. Thus, the characteristics of parents (especially the mother) and the hometown can confound the relationship between SES and only-child status, so we consider these factors in the following analyses.

## 5. Conceptual framework

This section presents a theoretical model that relates the two outcomes estimated in the following sections. Estimand in this study are the average values of and For the former, we look at the likelihood that an only child marries a person of type $p$, where $p \in \{OnlyChild, NonOnlyChild, Single\}$. On the other hand, the latter shows that only children's partners are likely to be more or less educated than non-only children's partners. Although each of the two outcomes provides information in capturing the characteristics of marriage patterns for only children, they can be linked together when considered in a standard decision-making model of marriage-matching.

Consider a frictionless marriage matching model in which a man and a woman form a family in the marriage market. We assume a non-transferable utility so that there is no pure transfer device within the household. In the model, a marriage requires a joint agreement where individuals get married only if one and their partner agree. In line with the recent accumulated literature, our model allows multiple-dimensional attractiveness in marriage sorting (i.e., one-child status and other attractions) as in Pierre-André Chiappori, Oreffice, and Quintana-Domeque (2012) and Pierre-André Chiappori, Oreffice, and Quintana-Domeque (2018).

Specifically, we consider the marital surplus of each individual as follows.

$$S(s, s_p, x, x_p) = M(s, s_p) + U(x_p) - V(s, x),$$

where $M(s, s_p)$ is the utility at marriage which depend on one's own and the partner's sibling traits $s, s_p$. The second term $U(x_p)$ is a utility depending on patner's other characteristics $x_p$. Finally, the third term $V(s, x)$ is an outside option depending on own sibling traits and other characteristics, $s, x$.

Using the property of marriage surplus, we can define the threshold of the partner's attractiveness for marriage $\bar{U}(x_p; s, s_p)$ as

$$\bar{U}(x_p; s, s_p) \equiv V(s, x) - M(s, s_p).$$

The equation above means that a person with the traits of $s, x$ prefers to marry the partner if and only if his other attractiveness exceeds the threshold $U(x_p; s, s_p) > \bar{U}(x_p; s, s_p)$.

We then assume the relation of utility at marriage. One is the case where the utility at marriage is smaller than that with a non-only child (i.e., $M(1, s_p) < M(0, s_p)$).[14] This

---

[14] Alternatively, we can consider the economy where the utility at marriage with only children is larger than that with a non-only child (i.e., $M(1, s_p) > M(0, s_p)$). In such an

applies the society where the economies of scale or a dynastic model to sustain the blood or legacy crucially matter for the marriage with only children (negatively), which is quite plausible from the argument of previous sociological literature (Yu and Hertog (2018); Uchikoshi, Raymo, and Yoda (2023)). Under this circumstance, we can have the following hypothesis showing the relations between two outcomes of marriage patterns. For proof of the derivation of the hypothesis, see Appendix B.

**Hypothesis.** *Suppose that $M(1, s_p) < M(0, s_p)$. Then, when the observed pattern $\{1, s_p\}$ is larger than $\{0, s_p\}$, $\bar{U}(x_p;\ 1, s_p) < \bar{U}(x_p;\ 0, s_p)$.*

Intuitively, if the utility at marriage with an only child is smaller than that with a non-only child for any type, the type $p$'s higher likelihood of marriage with an only child than that with a non-only child implies the lower reservation utility for type $p$'s other attractiveness than their only-child status (i.e., $\bar{U}(x_p;\ 1, s_p) < \bar{U}(x_p;\ 0, s_p)$). Consequently, the only children compromise on the partner's other attractiveness, $x_p$. Note that the unintuitive relation between the assumption of $M(1, s_p) < M(0, s_p)$ and the observed marriage patterns predicts a clear-cut direction of change in the partner's SES.

In the following analysis, we use SES (years of schooling) as a proxy of a partner's attractiveness other than their only-child status. Assume education level is a good proxy of $x$ or $x_p$. Then, we can rewrite the hypothesis for more specific empirical predictions.

**Empirical prediction.** *Suppose the utility at marriage with an only child is lower than that with a non-only child. When the probability of marriage with type-p partner is larger for only children than non-only children, only children are subject to a marriage matching penalty if they marry a type-p person.*

As discussed above, we look at the estimand, the average values of () and () in the following sections. In the next section, we examine the types of marriage partners for only children. Then, based on observed marriage patterns between only and non-only children, we test the hypothesis on premium or penalty by estimating the gap in partners' educational backgrounds. If we observe a relationship between the two outcomes consistent with the theoretical results, we can also predict the magnitude relation between utility at single of only children and that of non-only children (i.e., confirm whether only children are at a disadvantage in the marriage market).

---

economy, the higher utility at the marriage with an only child can be interpreted by the higher bequests or other forms of stronger descendant altruism for only children. However, it is discussed in the Appendix B that the results obtained from this assumption are inconsistent with the actual findings from the data.

## 6. Marriage patterns on only-child status

In this section, we first investigate whether only children are likely to marry or remain single, and with whom only children are likely to marry regarding the only-child status. This analysis not only brings the specific empirical hypotheses but allows us to know whether people choose mates in a positive or a negative assortative manner regarding only-child status.

In theory, we can expect two outcomes of assortativity in the marriage market, characterized by positive and negative assortative matching patterns (Becker 1991). The former results from an equilibrium in which people marry partners who share the same characteristics, which generally include religion, culture, ethnic preferences, professional interests, educational background, and physical characteristics. By marrying a partner with similar traits, people can reduce the cost of communication and other conflicts. For example, suppose that the universal economic incentive of scale economy and monetary incentive matters for choosing a marital partner and first caring for one's parents crucially imposes a specific cost. In this case, non-only children, less likely to bear the burden, will be the first to marry partners with the same trait and exit the marriage market. Then, only children would be more likely to marry other only children who remain in the marriage market. If the effect of the economic benefit, such as a bequest for only children, dominates the former, only children first leave the marriage market, followed by non-only-child couples.

Negative assortative mating describes the opposite equilibrium, whereby individuals avoid marrying those similar to them. A possible cause of this pattern could be the benefit of trade between providing domestic public goods (i.e., caregiving and child-rearing) and economic resource. Suppose the benefits from this public good are sufficiently large for the couple. In this case, it is optimal to associate with those with different characteristics to trade according to the relative magnitude of opportunity costs within the household. For example, consider this issue in the marriage of a one-child individual. When this effect is large enough, an only child with a high need to look after her own parents avoids an only child with a high opportunity cost of caring for a partner's parents. Instead, she will partner with a non-only child with a low opportunity cost for looking after his own parents and a lower arrangement cost for caring for both parents. In addition to the arrangement of public good provision, the only children's higher economic resource of endowment and possibly the strong incentives for maintaining family continuity (i.e., dynastic model) enhance the benefit of trading between only and non-only children.

### 6.1 Estimation method

For the purpose, we examine the likelihood of each marriage pattern. We present estimates of how marital status differs between only children and non-only children, controlling for gender, year of birth, age, and birthplace. We use the augmented inverse probability weighting (Robins and Rotnitzky 1995; Chernozhukov et al. 2017). The estimator requires to estimate the conditional means of $Y$ and $OnlyChild$, $E[Y|OnlyChild, X]$ and $E[OnlyChild|X]$, where $Y = \{Married\ with\ OnlyChild, Married\ with\ NonOnlyChild, Single\}$ and $OnlyChild = 1$ if an individual is an only child. $X$ is control variables including own years of schooling, birth year, age, and region of birth. We implement the stacking algorithm (including OLS, random forest, and Bayesian additive regression trees) to estimate conditional means without parametric assumptions. Note that we use robust standard errors clustering household-level.

We estimate this equation for each male and female sample. Technically, it estimates the difference between the average share of people in the marital status (married with a $p$-type partner or remained single) in the only-child population and that in the non-only-child population. The coefficient, thus, shows how one's only-child status affects the likelihood associated with each marriage state. The values indicate the gap in the proportion of marriage states in the measure of a percentage point.

## 6.2 Results: Partner's type on only-child status

Figure 1 is around here

Figure 1 presents the estimated coefficients for $OnlyChild$ conditional on the type of marriage partner (as well as single status) and then sorted by gender. The left panels show the results for single, and the right panels show the results for the marriage with only children. Note that we only indicate the results of two status as the effects of rest group of those marrying with a non-only child are automatically calculated with these two results.

We can learn four things by comparing the coefficients across the panels. First, looking at the panel as a whole, we observe similar trends in each panel regardless of gender. Second, when we look at the panel of singles, only children are more likely to remain single than non-only children, and this trend is stronger among men. The coefficients for only children are higher at 0.05 for men and 0.02 for women. This gender difference may stem from cultural differences in the roles of husband and wife in two families united by marriage, as pointed out by Yu and Hertog (2018). In Japan, women are used to joining the groom's family upon marriage, and the effect of this assumption is more significant in the case of an only child. If a man's family is given priority even if he marries an only-child woman, the man may not mind if his partner is an only child. However, if married to an only-child man, his wife is likelier to join the husband's family and bear its various obligations. In such a social context, women may avoid marrying only-child men. As a result, only-child men may be more likely to remain single. Third, only children are more likely to marry only-child partners. The values are now 0.07 for men and 0.08 for women. Finally, inextricably associated with the results above, the only-child status reduces the likelihood of marrying a non-only child partner, at -0.12 for men and -0.1 for women.

This analysis suggested two important points further. First is that people choose mates in a positive assortative manner regarding only-child status.[15] Second, we can present more specific empirical hypotheses by combining the theoretical results in the previous section and the difference in the observed likelihood of marriages together. Suppose the utility at marriage with only children is smaller than that with non-only children. Then we observed that the likelihood of marriage with only children is larger for only children. In this case, we predict that the only children compromise on the partner's attractiveness when they marry an only child (i.e., only children's partners are less educated than non-only children's partners when conditioning the partners of both groups to the only children). Recall that we refer to the lower (higher) educational background of only children's partners as the only-child matching penalty (premium) and measure them according to the hypothesis in the next section.

## 7. Only-child matching premium/penalty in the marriage market

The analyses in the previous section have shown a trend of positive assortative mating with a higher likelihood of marriage between two only-child individuals than between an only child and a non-only child. When the surplus of marriages with only children is smaller than that with non-only children, only-child individuals may compromise on their partner's attractiveness in the marriage with only-children partners, so we expect that an only-child matching penalty exists in the marriage market.

Ideally, we would like to measure the only-child matching premium/penalty defined on a utility basis. However, this is impossible because it is unobservable. Thus, we alternatively use the partner's socioeconomic status (years of schooling) as its approximation. Human capital is likely a monotonic form of attractiveness, and it is practical to link with the arguments on inequality among households (Mare 1991; Pencavel 1998; Fernández and Rogerson 2001; Breen and Salazar 2011; Greenwood et al. 2014, 2016; Eika, Mogstad, and Zafar 2019). We here apply Pierre-André Chiappori, Oreffice, and Quintana-Domeque (2018) to estimate the matching premium/penalty regarding partner SES. It should be noted that higher SES may not necessarily account for women's attractiveness, unlike men's (Fisman et al. 2006; Hitsch, Hortaçsu, and Ariely 2010; Low 2014; Bertrand, Kamenica, and Pan 2015). Therefore, the only-child premium (penalty) in terms of higher (lower) partner SES may be more sensitive among women.

_______________________

[15] Although we here limit ourselves to likelihood comparisons to intuitively understand the state of the marriage market, we conduct a formal analysis to measure assortativity on only-child status in Appendix C, following Pierre-Andre Chiappori, Costa Dias, and Meghir (2021). Using multiple other indices, we further confirm positive assortative mating on only-child status.

## 7.1 Estimation method

We regress the partner's years of schooling on the only-child status and other control variables as in the previous section. In this analysis, $Y$ isinstead the partner's SES (years of schooling), which captures other attractiveness than only-child status. Sharing with previous analysis, $OnlyChild$, our variable of interest, is a dummy variable that takes one for an only-child individual. $X$ includes own years of schooling, birth year, age, and region of birth. Again, we control for the individual's own years of education to focus on the disparity in adulthood and to consider positive assortativity on education for couples.

## 7.2 Results: Partner's years of schooling (Pooled sample analysis)

Figure 2 is around here

Figure 2 shows the results for the coefficients of $OnlyChild$ using the total sample by gender. The panel demonstrates the coefficient of only-child status on the spouse's education. Figure 2 does not show a significant only-child matching premium or penalty in the pooled sample for both men and women.

## 7.3 Results: Partner's years of schooling (Subsample analysis by partner's status)

As we have seen, the difference is small and unclear in the results of the pooled data. However, the partner's only-child status may matter in determining the premium/penalty. Recall the hypothesis depending on the relative size of the surpluses with only children or those with non-only children. The theory and the observed marriage patterns predict an only-child matching penalty for marriage with an only child. Therefore, in this subsection, we analyze the SES penalty using subsamples characterized by the partner's only-child status.

Figure 3 is around here.

Figure 3 shows the results of a subsample analysis restricted to gender and the only-child status of the marriage partner. The figure is similar to Figure 2 in the pooled sample analysis, but with the upper panels restricted to the male sample and the lower panels restricted to the female sample. Furthermore, we demonstrate the results in the left panels with individuals whose marriage partner is a non-only child and in the right panels with individuals whose marriage partner is an only child.

Figure 3 shows that, as discussed in Section 5, the results differ significantly depending on the marriage partner. In the sample with a non–only-child spouse, the coefficients for both men and women are closer to zero, implying that one's only-child status does not affect the partner's attractiveness in terms of educational background. However, when restricted to the sample where the partner is an only child, the coefficients move away from the 0 lines. The value is -0.03 for men (not significant at 5%) but negative -0.57 for women and significant at the 5% level. The coefficient size of female only-child status with an only-child partner is comparable with the magnitude of the gender gap in education of our sample at 0.54. In summary, we observed two marriage outcomes for only children consistent with our canonical model: the probability of marriage with an only child is high for only children, and they suffer a marriage matching penalty when marrying these only children. These two empirical findings imply that only children face disadvantages in the marriage market, and this is derived from a theoretical model suggesting that they have sufficiently lower utility in remaining single (see Appendix B). In line with the findings from online dating studies (Yu and Hertog (2018)), we have demonstrated that only children face a disadvantage in the marriage market, even when it comes to real marriages.

# 8. Supplementary analyses

Thus far, we have observed that only-child individuals incur penalties in terms of lower partner SES. In addition to their less benefit from scale economies, only-child individuals incur a matching penalty, especially when they marry an only-child partner, which may cause more considerable inequality. In this section, we then attempt to deepen our understanding of the effects on marital outcomes of only children from two perspectives. One is to conduct a heterogeneity analysis, and the other is to analyze based on alternative sibling configurations.

## 8.1 Heterogeneity of only child penalty

In this subsection, we examine how the one-child matching penalty is affected by heterogeneity. In particular, we focus on two heterogeneities. The first is the birth year. The meaning of an only child may change over time. The second is educational background. A higher level of education may increase one's attractiveness and may weaken the effect of being an only child. In addition, if caregiving duty is critical for their attractiveness, one may be able to purchase the services in the market if they can sufficiently afford it. To test these issues, we examine the effects of heterogeneity separately for men and women using Semenova and Chernozhukov (2021)'s method.

Figure 4 is around here.

Figure 4 is the result of the heterogeneity analysis of the marriage patterns. Let us first look at the effect of the birth year. We see that the more recently they were born, the stronger the tendency of the main result. Only children are more likely to remain single, less likely to marry non-only children, and more likely to marry only children. When we look at education, it is seen that the trend in the main result weakens. Now, consider only children from the view of economic incentives of scale economy and monetary transfer from parents for these results.

Figure 4 shows counterintuitive results for birth year if we take account of the circumstance where Japan has experienced the gradual socialization of elderly care through public policies (and slowly fading social norms on filial obligation).[16] However, this trend of socialization of care may also reflect the trend of women's advancement in society and the decrease in their labor in households. Women, who were typically used to caring for their parents and the parents-in-law, are now difficult to rely on as caregivers. Consequently, the difficulty of caregiving arrangements within household(s) may lead to the enhanced trend that people choose mates in a positive assortative manner regarding only-child status (i.e., only children are less likely to be chosen as a marital partner). In terms of another factor, monetary transfer, only children's attractiveness may be weakened by the declining relative value of the monetary transfers from parents than their own economic capabilities.

When analyzing the heterogeneity on age, it becomes evident that the main trend becomes more prominent as respondents get older at the time of the survey. From the fact that younger individuals to remain single, the result reflects the trend regarding the choice of marriage partner becomes even stronger when comparing those who are already married and young to those who are older. This suggests that individuals who marry at a later age may be selecting their partners with a more realistic understanding of caregiving responsibilities. The result on educational level may reflect that only children can overcome their disadvantages if one's educational background is higher, as already discussed.

Figure 5 is around here.

Finally, Figure 5 presents estimates of the matching premium/penalty measured by the partner's SES. From the analysis, it is shown that the main results are not significantly affected by either own birth year or education. In sum, heterogeneity affects only the choice of marital partner based on sibling structure but partner's SES.

---

[16] Historically, children's burden of filial duty has been declining. First, in 1961, Japan enacted a national pension system. Additionally, with the enactment of the Long-Term Care Insurance Law in 2000, a system was put in place for society to support the care of elderly individuals.

## 8.2 Excluding Education from Controls

This subsection tests whether our main result holds without controlling for education. As discussed in the main text, only children tend to be more educated, which may cancel out the penalty. It is also interesting to see how being an only child, rather than an only child and non-only child standing in the marriage market, affects the partner's education overall and its impact after considering the positive assortativity of education. Therefore, we followed the same procedure as in the main text and removed one's own education from the control variables. The sibling composition of the marriage partner is indicated by Figure 6, while the educational background of the marriage partner is indicated by Figure 7.

Figure 6 is around here.

Figure 7 is around here.

From the Figure 6, we can see that only children are more likely to remain single and more likely to marry an only child, unchanged from the main results without controlling for education. Also consistent with the main result is that the effect on singleness is more prominent for males. On the other hand, according to Figure 7, the results for years of education differ slightly from the main result. A clear difference was evident depending on the type of marriage partner. Only children married to a non-only child significantly increased their partner's education, although the coefficient was relatively small. However, the partner's education for an only child who is married to an only child significantly decreased for both males and females. As in the main results, this was particularly true for women, and the coefficients were about the same as in the main results.

In sum, although the impact is small, what differs from the main result is a premium for those married to non-only children. This trend, resulting from the lack of education control, is expected to reflect the significant positive correlation between education levels in partner selection. However, the results do not violate our theoretical prediction of a penalty for marrying an only child. Thus, our empirical results based on our theoretical predictions are robust even without controlling for education. Taken together with the results of the heterogeneity analysis in Section 8 (the less educated are more likely to marry an only child), it seems likely that those who are not well-educated face a more severe marriage matching penalty for being an only child.

## 8.3 The effects of alternative sibling positions

So far, we have examined the effect of being an only child on marriage patterns, but we still need to understand the underlying factors causing these effects. To gain further insights, we conducted an additional analysis considering alternative sibling positions and their interpretation in terms of intergenerational relationships. If these relationships are responsible for the disadvantages experienced by only children in the marriage market, policy interventions such as promoting the socialization of informal care could help reduce these penalties.

In this context, we explored two alternative sibling positions: the effect of being the eldest son and the effect of being the eldest child. In Japan, the eldest son (and his wife) traditionally bears certain obligations, including caring for his parents. Similarly, the eldest daughter with no male siblings was expected to assume this role. While the influence of the eldest son is widely known, the concept of primogeniture, where inheritance goes to the eldest child, suggests that birth order might be more significant than simply being the eldest son. Recent trends indicate a growing preference for individuals to take care of their own parents, and they also expect their own children to care for them. Additionally, considering the persistent gender gap in the provision of informal care, women may be expected to care for their parents even if they have younger brothers. If there is a significant age difference between the two siblings, the first-born effect may be even more pronounced than that of the eldest son. Taking these two cases into account, we tested whether the effects of these sibling positions exist. To focus on sibling positions, we analyzed samples from families with only two siblings. This approach allowed us to extend our main findings on the only-child effect and make meaningful comparisons.

In our analysis, one's position could be determined by the position of the other sibling. For instance, if one is a male and the first-born child, he is considered the heir if the other sibling is a younger brother, an elder sister, or a younger sister. If one is a female, the first-born daughter without a male sibling is considered the heiress. Thus, we defined a dummy variable called "Patrilineal" that takes a value of one for males when the other sibling is a younger brother, elder sister, or younger sister, and zero if he has an older brother. For females, the dummy variable takes a value of one if the other sibling is a younger sister, and zero if the other sibling is an elder brother, younger brother, or elder sister. On the other hand, if we focus on birth order, both males and females become heirs if they are the first-born children. Hence, we defined a dummy variable called "Primogeniture" that takes a value of one when the other sibling is a younger brother or sister, and zero if they have an older brother or sister. Finally, we categorized individuals' marital status into three groups: married with an only child, married with a non-only child, and remaining single. This allowed us to compare the results with our main findings.

Figure 8 is around here.

Figure 9 is around here.

Figure 8 shows the results of the effects of heirs on their marital statuses in the two definition. The results indicate that there are no significant differences in terms of marital status when considering the Patrilineal Heir variable. On the other hand, it is observed that Primogeniture heirs are less likely to remain single. When analyzing the education level of the marriage partner as in Figure 9, no significant effects are observed for either definition. However, in the case of Patrilineal Heir, the coefficient is larger for females when the marriage partner was an only child, which aligns with the main finding. Overall, the penalties associated with each dummy variable are not as pronounced as the effect observed for individuals who were only children.

In sum, the overall results are partially consistent with the effects observed for only children when considering the dummy variables representing the respective heirs. These findings do not dismiss the possibility that the strength of intergenerational relationships plays a role in the penalties faced by only children. Furthermore, the penalty for being an only child is greater than that for being an heir with siblings, as suggested by the two definitions. This highlights the potential challenges faced by only children, who lack the support of siblings, in contrast to heirs whose responsibilities can be shared among siblings. However, it is important to note that this study did not establish formal causal effects, so not all of the penalties can be attributed solely to generational relationships.

## 9. Conclusion

This study explored marriage matching among only-child individuals, focusing on their solid intergenerational relationships. Specifically, we first examined the likelihood of marriage experienced by an only-child individual by partner type to see what marriage patterns emerged. The analysis revealed that the likelihood of staying single is higher for only children. Moreover, being an only child increases the likelihood of marrying another only child but reduces the likelihood of marrying a non-only child. This result of positive assortative mating on only-child status predicts that only children compromise on the attractiveness of their marriage partners (i.e., suffer a matching penalty).

To measure this matching penalty, we applied Pierre-André Chiappori, Oreffice, and Quintana-Domeque (2018) framework to our analysis, finding an only-child matching penalty in terms of lower partner SES. Furthermore, we observed that this penalty was more pronounced for couples with two only children, as predicted by the finding of positive assortative mating on only-child status. Thus, the observed only children's marriage pattern and their matching penalty are consistent with our theoretical model.

Moreover, we conducted additional analyses to gain a more profound understanding of the underlying cause of the penalty. Heterogeneity analyses revealed that one's own educational level helps alleviate the disparity in partner choice. Additionally, assortativity based on the only-child status becomes more pronounced among respondents born more recently and those who are older. Moreover, other analyses exploring alternative sibling positions do not refute the possibility that heavier filial obligations influence the marriage patterns of only children.

Several conclusions can be drawn from these findings. Firstly, the composition of siblings, which is an inherent factor beyond an individual's control, tends to disadvantage some individuals in terms of finding a suitable match in the marriage market. Specifically, the results indicate that being an only child led respondents to compromise on their partners' socioeconomic status (SES) attractiveness in their marriages. Secondly, the marriage market exacerbates disparities in family sizes. Considering that only children bear a heavier responsibility in caregiving, our results indicating that only children are more likely to remain unmarried or marry other only children imply an increase in the inequality of caregiving burdens through the marriage market. Regarding the penalties faced by only children, socializing the burden of care may be expected to address two negative aspects: the disadvantageous marriages of only children and the widening gap in caregiving burdens among the younger generation.

Before closing our study, we discuss the limitations of this study and future research directions. The only information that we have on sibling composition is on *surviving* siblings. Since we include in the sample those that have already lost siblings, we may underestimate the penalty for older generations. The issue also fails us to consider the possibility that only-child status may be driven by biological factors (e.g., infertility, low probability of survival of all siblings), as pointed out by Lu and Vogl (2022). Since families with weak constitutions may also be at a disadvantage in the marriage market, data that include this information would allow other possible interpretations.

Relatedly, this study used data from Japan, one of the East Asian countries with the strongest traditions of filial piety, to focus on intergenerational relationships where policy intervention is possible. While the results were somewhat reasonable, other sources of explanation for the only-child penalty are possible, as discussed above. It would be interesting to see the effects of variation in policy changes on the marriage patterns, if any, in other economies a la Bau (2021). She demonstrates that pension policies implemented in societies dominated by both males and females have had an impact on cultural changes for marriage customs. This study can also be expanded to analyze the impact of alternative sibling structures on marital outcomes. While the current analysis primarily concentrated on the only-child status, which is less influenced by specific periods and cultures, a potential avenue for future research is to comprehensively investigate the effects of sibling structure in conjunction with those of masculine culture.

Finally, we believe that the findings in this study will provide valuable insights into the speed of population decline. Vogl (2020), in their research on the evolutionary process of intergenerational associations in fertility, has raised the possibility of marriage assortativity as a mechanism that may contribute to this acceleration. Our discovery of assortativity represents a significant step forward in our understanding of demographics. It would be meaningful to examine the demographic impact of positive assortativity on sibship size in the marriage market, as it could potentially contribute to the further decline in fertility rates.

## Table

*Table 1: Descriptive statistics of respondents' characteristics by only-child status*

| Characteristic | **0**, N = 43,870 | **1**, N = 3,111 |
|---|---|---|
| **Gender** | | |
| Men | 21,963 (50%) | 1,557 (50%) |
| Women | 21,907 (50%) | 1,554 (50%) |
| **MarriageStatus** | | |
| Married | 39,270 (90%) | 2,663 (86%) |
| Single | 4,600 (10%) | 448 (14%) |
| **Age** | 47 (37, 56) | 46 (37, 55) |
| **Years of Schooling** | 12.00 (12.00, 14.00) | 14.00 (12.00, 16.00) |
| **Birth Year** | 1,955 (1,946, 1,965) | 1,957 (1,948, 1,966) |
| **BirthPlace** | | |
| Chugoku | 2,817 (6.4%) | 212 (6.8%) |
| Hokaido | 2,317 (5.3%) | 199 (6.4%) |
| Hokuriku | 2,481 (5.7%) | 175 (5.6%) |
| Kinki | 6,171 (14%) | 535 (17%) |
| Kyusyu & Okinawa | 6,783 (15%) | 387 (12%) |
| NorthKanto & Koushin | 4,068 (9.3%) | 230 (7.4%) |
| Shikoku | 1,888 (4.3%) | 115 (3.7%) |
| SouthKanto | 8,471 (19%) | 723 (23%) |
| Tohoku | 4,057 (9.2%) | 209 (6.7%) |
| Tokai | 4,817 (11%) | 326 (10%) |

*Table 2: Descriptive statistics of partners' characteristics by respondents' only-child status*

| Characteristic | **Not Only Child**, N = 39,270 | **Only Child**, N = 2,663 |
|---|---|---|
| **Only child status** | | |
| Not Only Child | 36,972 (94%) | 2,263 (85%) |
| Only Child | 2,298 (5.9%) | 400 (15%) |
| **Years of Schooling** | 12.00 (12.00, 14.00) | 12.00 (12.00, 16.00) |

## Figures

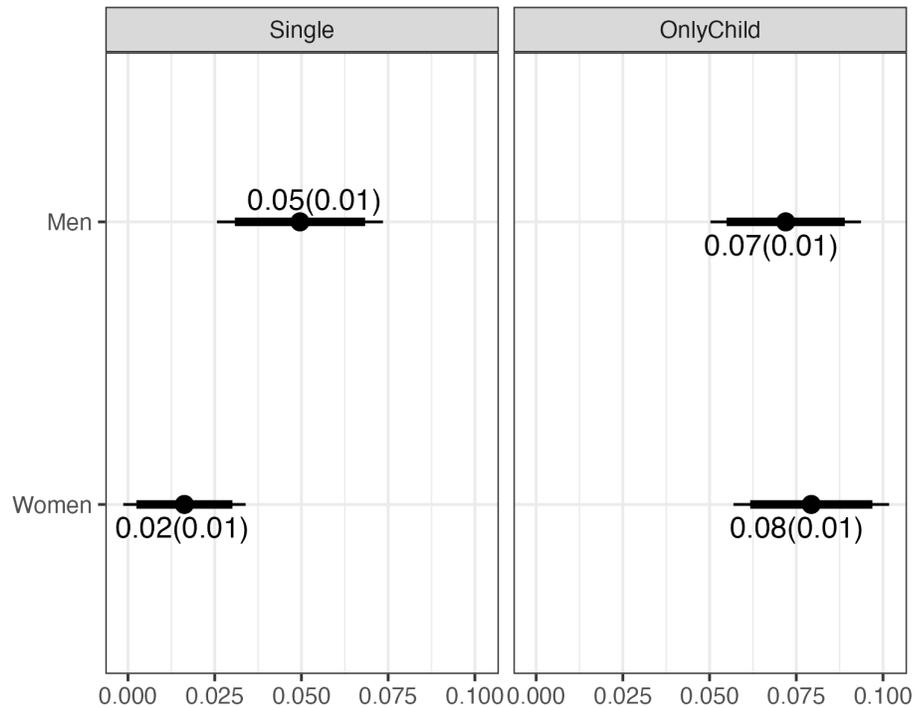

*Figure 1: Only-child status on partner's type*

Notes: This figure shows the different marital statuses according to only-child status: single (left graph) and married with only child (right graph), estimated by equation (1), along with the 95th confidence interval (bold lines show Bonferroni-corrected confidence intervals). All nuisance functions are estimated using the stacking learner, which consists of OLS (including squared terms of age, birth year, and education year), random forest, and Bayesian additive regression trees. We compare groups comprised only-child and not only-child.

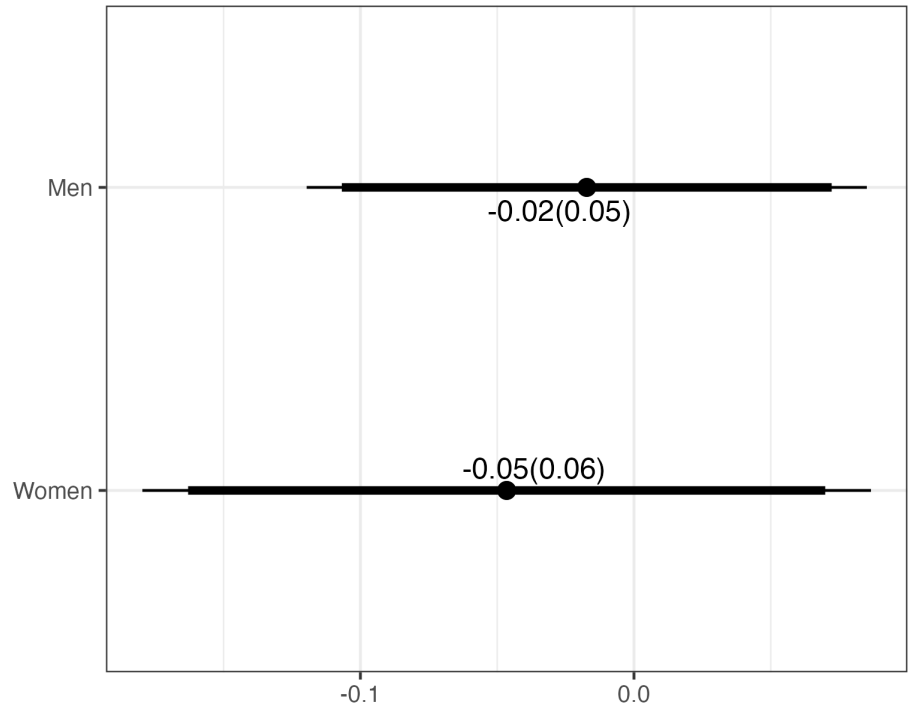

*Figure 2: Only-child status on partner's years of schooling (Pooled sample analysis)*

Notes: This figure shows the difference in partner's years of education according to only-child status, estimated by equation (2), along with the 95th confidence interval (bold lines show Bonferroni-corrected confidence intervals). All nuisance functions are estimated using the stacking learner, which consists of OLS (including squared terms of age, birth year, and education year), random forest, and Bayesian additive regression trees. We compare groups comprised only-child and not only-child.

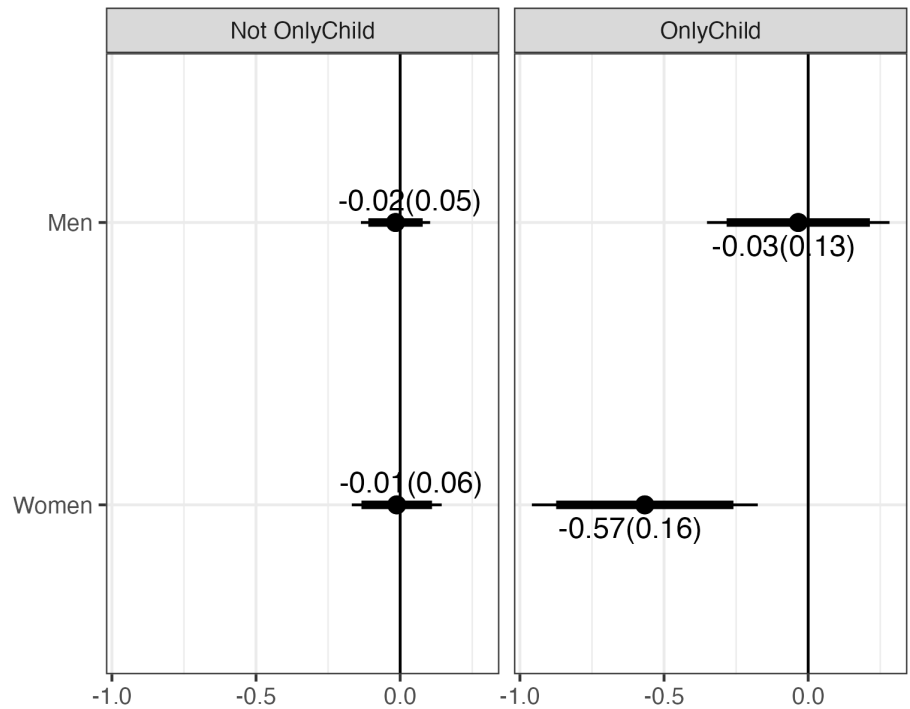

*Figure 3: Only-child status on partner's years of schooling (Subsample analysis)*

Notes: This figure shows the difference in partner's years of education according to only-child status by subsamples: married with not only child (left graph) and only child (right graph), estimated by equation (2), along with the 95th confidence interval (bold lines show Bonferroni-corrected confidence intervals). All nuisance functions are estimated using the stacking learner, which consists of OLS (including squared terms of age, birth year, and education level), random forest, and Bayesian additive regression trees. We compare groups comprised only-child and not only-child.

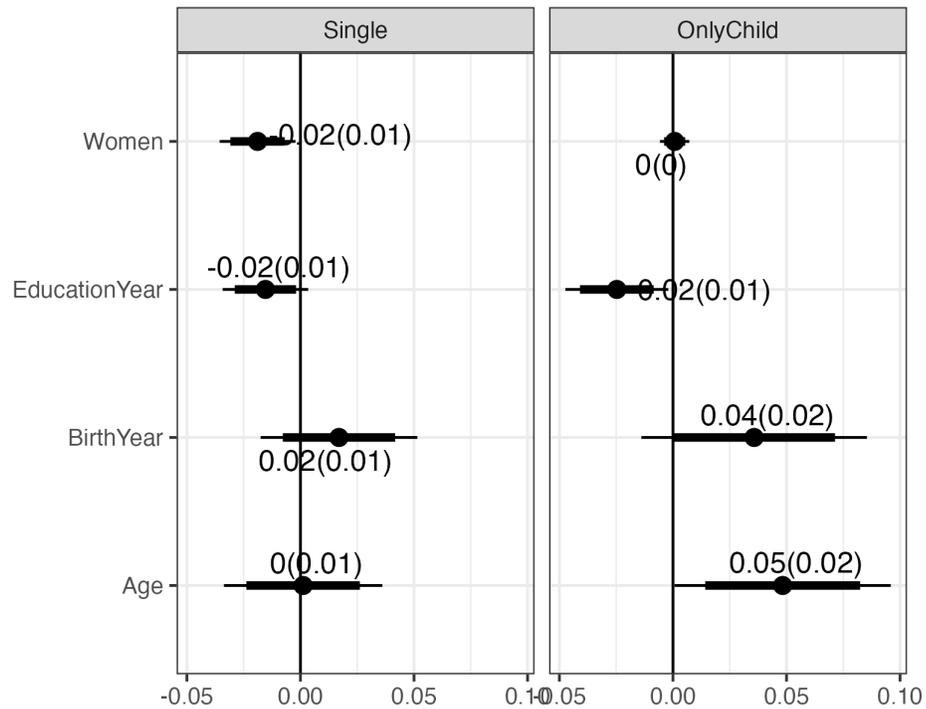

*Figure 4: Only-child status on partner's type (Heterogeneity)*

Notes: This figure shows the best linear projection of the conditional difference of marital statuses on only child status: single (left graph) and married with only child (right graph), along with 95th confidence interval (bold lines show Bonferroni-corrected confidence intervals). All nuisance functions are estimated using the stacking learner, which consists of OLS (including squared terms of age, birth year, and education level), random forest, and Bayesian additive regression trees. We compare groups comprised only-child and not only-child.

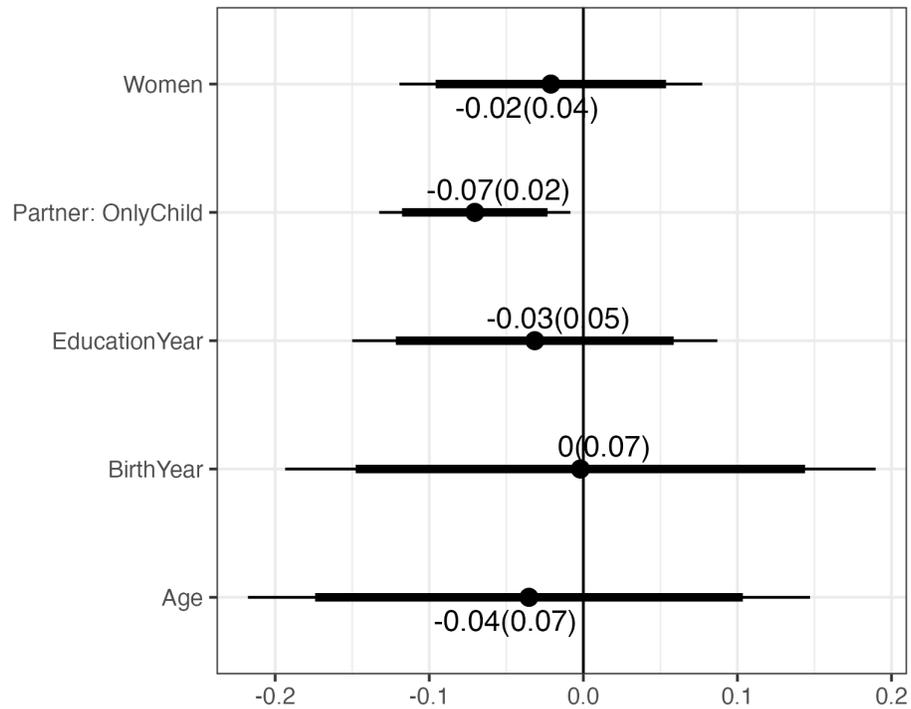

*Figure 5: Only-child status on partner's years of schooling (Heterogeneity)*

Notes: This figure shows the best linear projection of the conditional difference of the partner's years of education, along with the 95th confidence interval (bold lines show Bonferroni-corrected confidence intervals). All nuisance functions are estimated using the stacking learner, which consists of OLS (including squared terms of age, birth year, and education level), random forest, and Bayesian additive regression trees. We compare groups comprised only-child and not only-child.

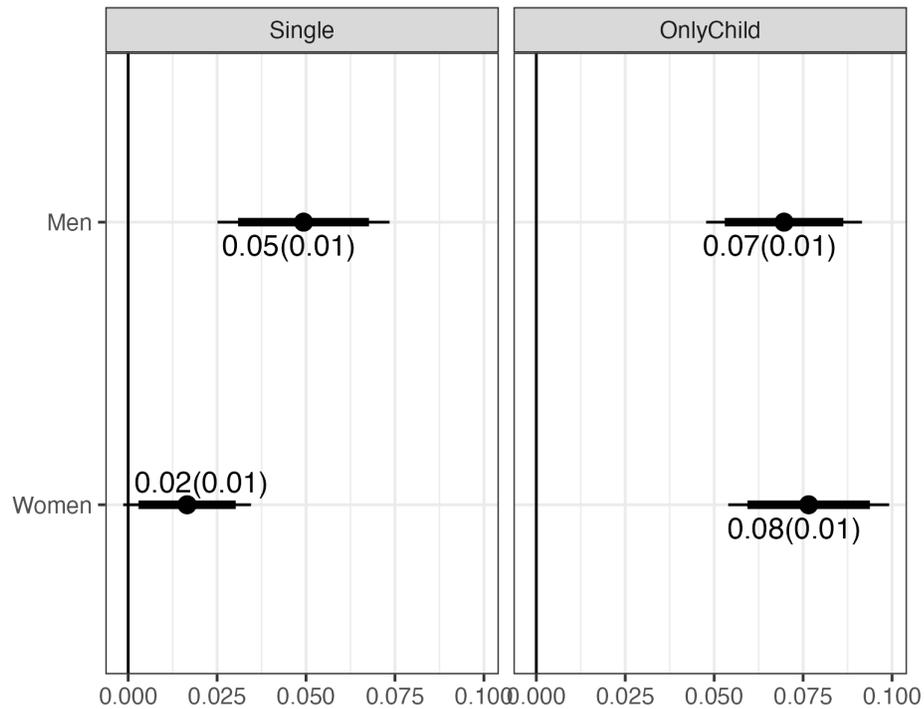

*Figure 6: Only-child status on partner's type (Without controlling own years of education)*

Notes: This figure shows the different marital statuses according to only-child status: single (left graph) and married with only child (right graph), estimated by equation (1) excluding own years of schooling from control variables, along with the 95th confidence interval (bold lines show Bonferroni-corrected confidence intervals). All nuisance functions are estimated using the stacking learner, which consists of OLS (including squared terms of age, birth year, and education year), random forest, and Bayesian additive regression trees. We compare groups comprised only-child and not only-child.

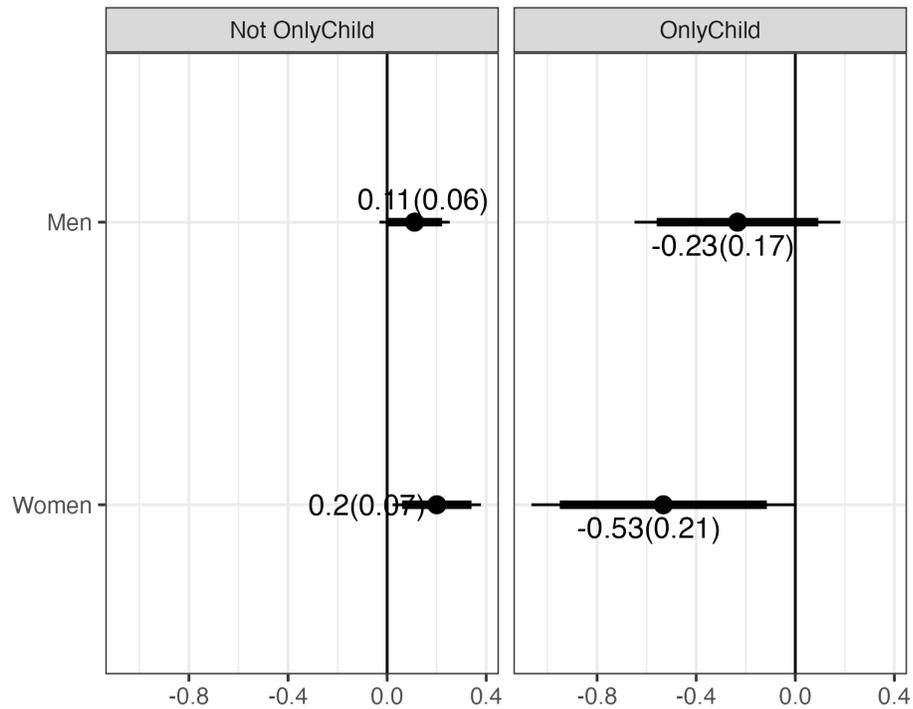

*Figure 7: Only-child status on partner's years of schooling (Without controlling own years of education)*

Notes: This figure shows the difference in partner's years of education according to only-child status by subsamples: married with not only child (left graph) and only child (right graph), estimated by equation (2) excluding own years of schooling from control variables, along with the 95th confidence interval (bold lines show Bonferroni-corrected confidence intervals). All nuisance functions are estimated using the stacking learner, which consists of OLS (including squared terms of age, birth year, and education year), random forest, and Bayesian additive regression trees. We compare groups comprised only-child and not only-child.

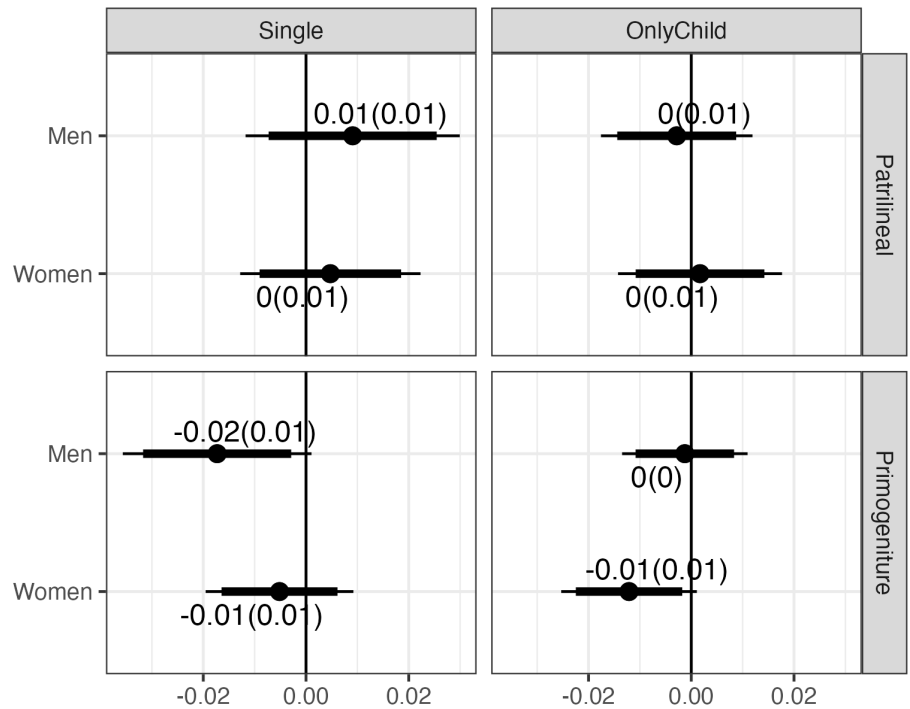

*Figure 8: Only-child status on partner's type (Alternative treatment)*

Notes: This figure shows the different marital statuses according to sibling positions: single (left graph) and married with an only child (right graph), along with the 95th confidence interval (bold lines show Bonferroni-corrected confidence intervals). The sibling positions are defined by dummy variables called "Patrilineal" (top graph) and "Primogeniture" (bottom graph) among two sibling respondents. Patrilineal takes a value of one for males when the other sibling is a younger brother, elder sister, or younger sister, and zero if he has an older brother. For females, the dummy variable takes a value of one if the other sibling is a younger sister and zero if the other sibling is an elder brother, younger brother, or elder sister. Primogeniture takes a value of one when the other sibling is a younger brother or sister and a value of zero if they have an older brother or sister for both males and females. All nuisance functions are estimated using the stacking learner, which consists of OLS (including squared terms of age, birth year, and education level), random forest, and Bayesian additive regression trees. We compare groups comprised of heir and non-heir in each definition.

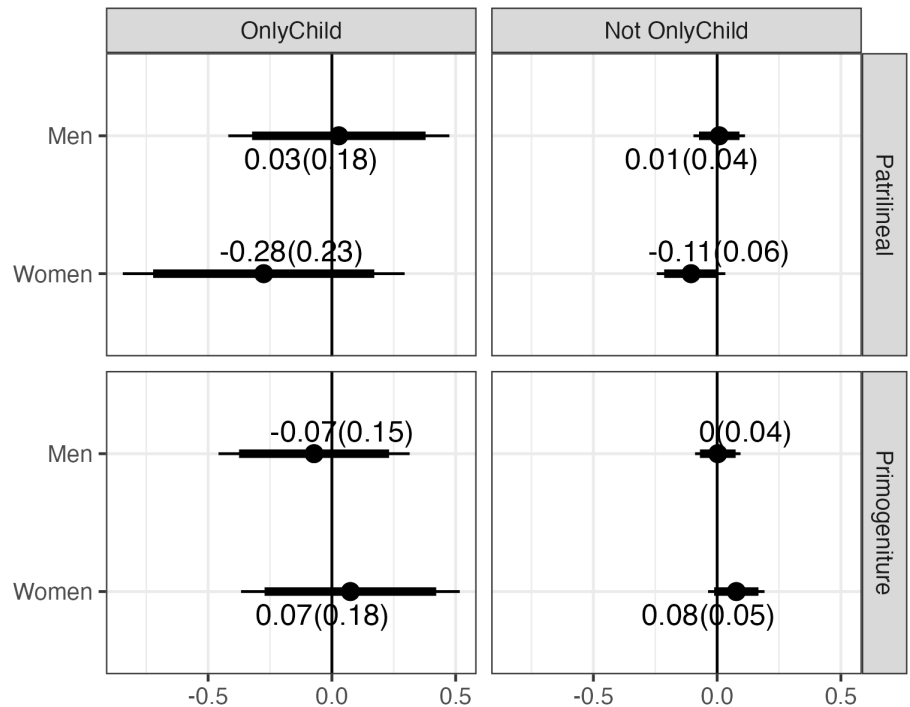

*Figure 9: Only-child status on partner's years of schooling (Alternative treatment)*

Notes: This figure shows the difference in partner's years of education according to alternative sibling position by subsamples: the results among the marriages with a non-only child partner (left graph) and the marriages with an only-child partner (right graph), along with the 95th confidence interval (bold lines show Bonferroni-corrected confidence intervals). The sibling positions are defined by dummy variables called "Patrilineal" (top graph) and "Primogeniture" (bottom graph) among two sibling respondents. Patrilineal takes a value of one for males when the other sibling is a younger brother, elder sister, or younger sister, and zero if he has an older brother. For females, the dummy variable takes a value of one if the other sibling is a younger sister and zero if the other sibling is an elder brother, younger brother, or elder sister. Primogeniture takes a value of one when the other sibling is a younger brother or sister and a value of zero if they have an older brother or sister for both males and females. All nuisance functions are estimated using the stacking learner, which consists of OLS (including squared terms of age, birth year, and education year), random forest, and Bayesian additive regression trees. We compare groups comprised of heir and non-heir in each definition.

# Appendix A. Previous litereture

In this appendix, we look at studies on educational outcomes, which is also our key variable, and then at studies on other factors that sibling composition may impact and thereby affect the lifetime welfare of the individual.

## Studies on sibling composition and educational outcomes

In this study, we measure only children's quality of marriage by schooling years, one of the dimensions of SES of the marriage partner. The individual's educational level may be related to the reason that he or she is an only child. It is also possible that the SES of the marriage partner, which is of interest, is significantly influenced by the SES of the individual through the assortative mating mechanism. In this study, we attempt to eliminate these effects by controlling for the individual's own education in the analysis. This subsection reviews previous studies examining the impact of an individual's sibling composition on educational outcomes to analyze the relationship between only-child status and the individual's educational background.

An impact of only-child status on socioeconomic outcomes that can in turn influence only children's marriage matching can be argued based on the idea of a quality–quantity trade-off in economics. Since Becker and Lewis (1973) and Becker and Tomes (1976) proposed a negative relationship, there has been theoretical development of this idea (Galor and Weil 2000; Hazan and Berdugo 2002; Moav 2005). The studies have further attempted to clarify causality by exploiting various strategies, including exogenous fertility increases due to multiple births (Rosenzweig and Wolpin 1980; Black, Devereux, and Salvanes 2005; Cáceres-Delpiano 2006; H. Li, Zhang, and Zhu 2008; Rosenzweig and Zhang 2009; Angrist, Lavy, and Schlosser 2010), variations in the gender composition of firstborns (Lee 2008) and first- and secondborns (Conley and Glauber 2006; Angrist, Lavy, and Schlosser 2010; Millimet and Wang 2011), variations in the severity of the OCP (Qian 2009; Rosenzweig and Zhang 2009; Liu 2014; B. Li and Zhang 2017; Qin, Zhuang, and Yang 2017), to find both a negative relation and no or even a positive relation between the two. Black, Devereux, and Salvanes (2005) also demonstrate a nonlinear relationship between sibship size and quality with Swedish panel data, controlling for family background and using the effect of twins on family size.

While not all studies are limited to only children, results from studies that replace the focus on two or more children with one on only children are also inconclusive. For example, Black, Devereux, and Salvanes (2005) and Qian (2009) show that only children are disadvantaged in their educational outcomes. However, Lee (2008) shows that only children have higher expenditures per child than siblings. Rosenzweig and Zhang (2009) find that being in a family with two twin children rather than one child reduces educational outcomes. Liu (2014) and B. Li and Zhang (2017) observe a generally negative effect of sibsize using variation in OCP severity as an instrumental variable. Qin, Zhuang, and Yang (2017) use regression discontinuity by the month of giving birth back-calculated from the time of OCP implementation and show a negative effect of an extra child on quality. Therefore, it is not possible to determine whether only children are more or less educated.

## Studies on sibling composition and other outcomes

In this subsection, we review studies that examine how sibling composition, including only-child status, affects individuals' welfare throughout life events. However, it should be noted that although these factors may all affect marriage matching, which is our outcome of interest, we have not been able to control for them in the study.

One can also consider the impact of being an only child on marital relationships from a developmental psychology perspective. Sibling relationships share many characteristics with romantic relationships. Because sibling relationships are long term and informal, those with siblings may find it easier to find a partner and maintain the relationship (Reese-Weber and Kahn 2005; Feinberg, Solmeyer, and McHale 2012). However, the findings on the psychological development consequences of only-child status are still inconclusive due to its complexity.

Historically, only children have been considered unique relative to children with siblings. Indeed, the childhood experiences of only children and children with siblings consistently differ across families. For example, in contrast to children with siblings, only children do not have to compete with others for parental attention or access to economic resources (Polit and Falbo 1987). In addition, only children have no social interaction with siblings in the family environment (Polit and Falbo 1987; Mancillas 2006) and have less experience with compromising with siblings and other peers.

Early research predicted that these differences in experience would negatively impact an only child's personality, making them selfish, spoiled, egotistical, and even narcissistic. A pioneering psychologist G. Stanley Hall once emphasized that "being an only child is a disease in itself" based on a study with an extremely small sample size of only children (as cited in Fenton (1928)), and this stereotype still permeates society (Mancillas 2006; Griffiths et al. 2021). However, recent studies have overturned this negative image, and numerous positive aspects of being an only child have been reported in academic literature, such as high creativity, resilience and success in adulthood (Blake 1989; Mellor 1990; Polit, Nuttall, and Nuttall 1980; Polit and Falbo 1987, 1988; Poston Jr and Falbo 1990).

Several studies have examined the relationship between sibling composition and labor market outcomes, treating them as child quality outcomes. Kessler (1991) showed that compared to middle children in families with three or more siblings, only children had lower employment rates between ages 14 and 22 for women but increased rates between ages 22 and 30. Black, Devereux, and Salvanes (2005) demonstrate that sibling structure matters not only for educational attainment but also earnings and full-time employment status in adulthood. The analysis finds that the effect of sibling structure is more prominent for women, with women born later (that is, having more siblings relative to only children) having lower incomes and being less likely to work full time. For men, being born later does not affect their full-time employment status but decreases their full-time income. On the other hand, Angrist, Lavy, and Schlosser (2010) show that an increase in the number of siblings does not have a clear impact on labor market outcomes. Despite the lack of consensus in the results on labor market outcomes, socioeconomic success itself significantly impacts marriage decisions. If only-child status matters for socioeconomic success, then socioeconomic variables are confounding factors for only children's marriage outcomes.

An enormous number of studies have examined which children care for aging parents in Japan and elsewhere (Konrad et al. (2002b); Whiteman, McHale, and Crouter (2003); Suitor and Pillemer (2007) among others). Many studies have shown that only children have a greater burden of caregiving than children with siblings (Coward and Dwyer (1990), Dwyer and Coward (1991), and Spitze and Logan (1991) for the US; Rainer and Siedler (2012) for European countries). Economic theory predicts from the geographical perspective that those with siblings try to leave their parents due to incentives to free-ride on other siblings (Konrad et al. 2002a; Rainer and Siedler 2009). Using a German micro dataset, Rainer and Siedler (2009) further show that this difference not only makes it harder for an only child to leave his or her parents but also results in lost labor market opportunities.

## Appendix B. Conceptual Model

In this Appendix, we derive our empirical hypothesis based on a simple theoretical model of marriage decision-making. We assume a non-transferable-utility model. The marriage surpluses of own and partner (indexed by $p$) are

$$S(s, s_p, x, x_p) = M(s, s_p) + U(x_p) - V(s, x)$$

and

$$S_p(s, s_p, x, x_p) = M_p(s, s_p) + U_p(x) - V_p(s_p, x_p),$$

where $s$ and $x$ are sibling structures and other characteristics, respectively. $s = 1$ indicates one is an only child, and $s = 0$ otherwise. $s_p = 1$ and $x_p$ also behave similarity for the one's partner. The first term for the first equation is the utility of marriage determined by the sibling composition of one and one's marriage partner. The second term represents the utility determined by the partner's attractiveness other than only-child status. The third term is the utility of being single, determined by one's only-child status and other characteristics. Finally, the marriage surplus of the marriage partner is defined in the same way. In further analysis, we assume the following.

**Assumption:** $M_p(1, s_p) < M_p(0, s_p)$ and $M(1, s_p) < M(0, s_p)$.

The assumptions implied are as follows. The first equation indicates that whatever the sibling composition of the marriage partner, the utility of one's partner is lower if the one is an only child. The second equation shows that, given the marriage partner's only-child status, the utility of the marriage state is higher for a non-one-child than for an only child. The assumption is justified if we consider an individual is an only child and has a higher burden of caring for his or her parents and is more likely to share that burden with their spouse. In that case, this corresponds to a lower utility for that marriage partner. On the other hand, the second equation captures that the financial or psychological costs (albeit loving) reduce the loss of utility at marriage, which is due to having a higher care burden for one's parents in their marital lives.

They are married if and only if the marriage surplus is sufficiently large ($S \geq 0$ and $S_p \geq 0$). Equivalently, there are thresholds (i.e., reservation utility of marriage characterized by the partner's attractiveness) as

$$\bar{U}(x;\ s, s_p) \equiv V(s, x) - M(s, s_p),$$

and

$$U_p(x) \geq \bar{U}_p(x;\ s, s_p) \Leftrightarrow S_p(s, s_p, x, x_p) \geq 0,$$

where

$$\bar{U}_p(x;\ s, s_p) \equiv V_p(s_p, x_p) - M_p(s, s_p).$$

These thresholds represent the minimum utility conditions necessary for an individual to enter into marriage. Note that these conditions become more stringent when the utility level of remaining single $V_p(s_p, x_p)$ is higher, or the utility from marriage with the partner $M_p(s, s_p)$ is lower.

Now that we have everything set up for the arguments, we explore the marriage decisions from both sides (one's own and their partner's) along with the mating patterns in the marriage market. Let us first look at the decisions from the partner's side. Given the partner's characteristics, the following equation holds.

$$\bar{U}_p(x_p;\ s = 1, s_p) - \bar{U}_p(x_p;\ s = 0, s_p)$$
$$= M(s = 0, s_p) - M(s = 1, s_p) \geq 0.$$

**Lemma 1.** *Both only-child and not only child partner tend to prefer not only child than not only child partner.*

Then, we consider the marriage decision from one's (only children's) own side along with the probability of their marriage with a partner from a specific group. First, recall that Lemma 1 implies that a larger probability of only children's marriage with a partner from a specific group requires $\bar{U}(x_p; s = 1, s_p) - \bar{U}(x_p; s = 0, s_p) < 0$ (i.e., the lower reservation utility for only children for the marriage with these partners). This is because they are likely to marry a specific type of partner although only children are avoided as partners from any type of individuals. In such circumstances, only children must compromise on their partners' attractiveness other than their only-child status to marry them. In sum, we can derive the following proposition.

**Proposition 1.** *Suppose that $M(1, s_p) < M(0, s_p)$. Then, when the observed pattern $\{1, s_p\}$ is larger than $\{0, s_p\}$, $\bar{U}(x_p;\ 1, s_p) < \bar{U}(x_p;\ 0, s_p)$.*

In the main discussion, we have been assuming that the benefits of marrying an only child are smaller compared to marrying a non-only child. However, it is theoretically possible to consider the alternative case where the surplus of marrying an only child is larger. The rationale behind this is that only children are often seen as more desirable marital partners because they tend to receive more financial support from their parents, such as inter-vivo transfers and inheritances.

In this alternative scenario, we could hypothesize that there is a premium for only children regarding the educational level of their partners, especially when they marry non-only children.[17] However, since we do not observe such a premium in reality, it contradicts the assumption that the marriage surplus with only children is larger.

_________________________

[17] Alternatively, we can consider the economy where the marriage surplus with only children is larger than that with a non-only child (i.e., $M(1, s_p) > M(0, s_p)$). In such an economy, the higher surplus of the marriage with an only child can be interpreted by the higher bequests or other forms of stronger descendant altruism for only children. In that case, we can derive the following hypothesis: if $M(1, s_p) > M(0, s_p)$ and the observed pattern $\{1, s_p\}$ is smaller than $\{0, s_p\}$, then $\bar{U}(x_p)$ of $s = 1$ is large than $s = 0$. To obtain the intuition, suppose that the surplus of marriage with an only child is larger than that with a non-only child for any type, but we observe a lower likelihood of marriage with an only child than that with a non-only child. In this case, it is expected that the higher reservation utility for type p's other attractiveness than their only-child status, i.e., $\bar{U}(x_p)$ of $s = 1$ is higher than $s = 0$. Thus, the only children are pickier about the partner's other attractiveness, $x_p$. The specific empirical prediction can be rewritten as follows: if $M(1, s_p) > M(0, s_p)$ and observed pattern $\{1, s'\}$ is smaller than $\{0, s_p\}$, then $s = 1$'s marriage partner is more educated than $s = 0$    In this case, we predict that the only children are picky about the partner's attractiveness when they marry a non-only child (i.e., only children's partners are more educated

To develop a consistent theoretical model that links the observed outcomes of partner type and partner education, we need to assume at least that only children generally avoid marrying other only children. This means that for only children, the utility at marriage is smaller when marrying another only child compared to marrying a non-only child. This assumption can be justified by considering factors such as the advantages of family economies of scale that largely apply to only children or the influence of the dynastic model.

## Appendix C. Assortativeness

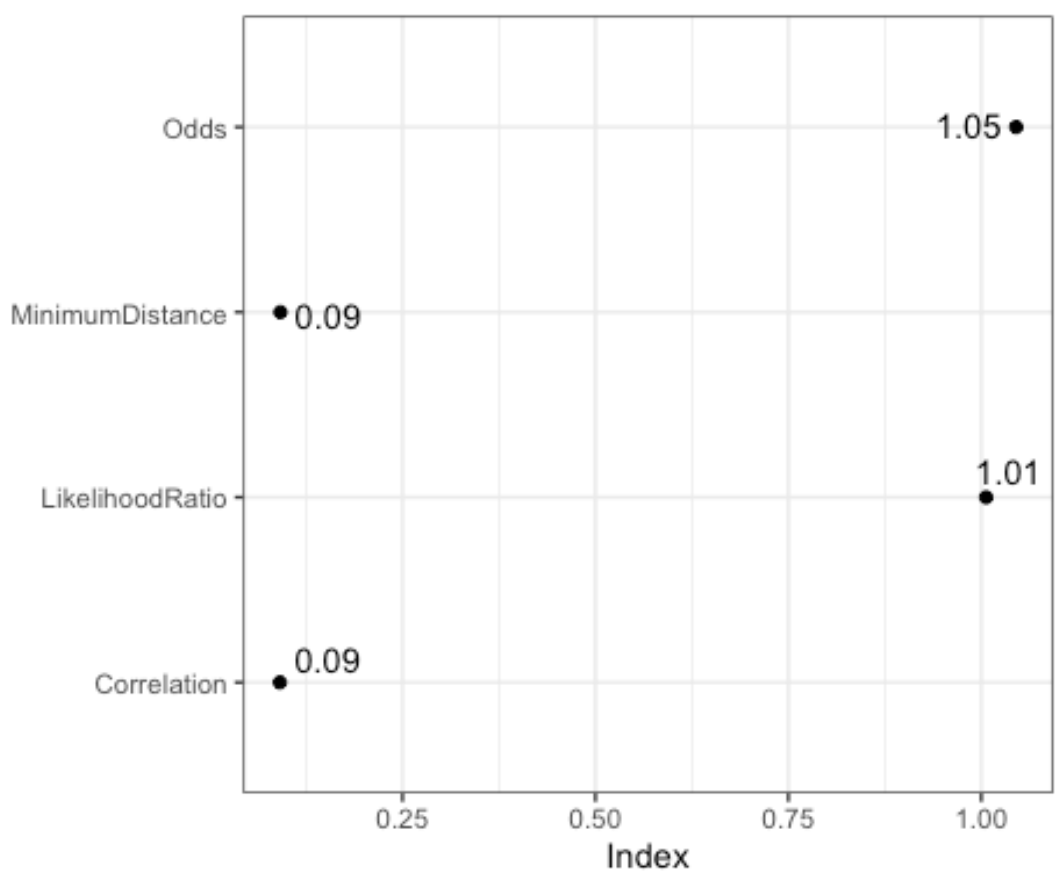

This appendix formally checks the assortativity of only-child status. We here use four indices presented in Pierre-Andre Chiappori, Costa Dias, and Meghir (2021). According to the definition of Pierre-Andre Chiappori, Costa Dias, and Meghir (2021),

───────────────────────────────

than non-only children's partners when conditioning the partners of both groups to the non-only children).

positive assortativity is observed in only-child status if the values for results of odds and likelihood ratio are greater than one, while it holds true if the values for results of the minimum distance and the correlation are positive. All indices show positive assortativity in the pooled data.

# References

Alesina, Alberto, and Paola Giuliano. 2010. "The Power of the Family." *Journal of Economic Growth* 15 (2): 93–125.

Angrist, Joshua, Victor Lavy, and Analia Schlosser. 2010. "Multiple Experiments for the Causal Link Between the Quantity and Quality of Children." *Journal of Labor Economics* 28 (4): 773–824.

Bau, Natalie. 2021. "Can Policy Change Culture? Government Pension Plans and Traditional Kinship Practices." *American Economic Review* 111 (6): 1880–1917.

Becker, Gary S. 1973. "A Theory of Marriage: Part i." *Journal of Political Economy* 81 (4): 813–46.

———. 1974. "A Theory of Marriage: Part II." *Journal of Political Economy* 82 (2, Part 2): S11–26.

———. 1991. *A Treatise on the Family: Enlarged Edition*. Harvard university press.

Becker, Gary S, and H Gregg Lewis. 1973. "On the Interaction Between the Quantity and Quality of Children." *Journal of Political Economy* 81 (2, Part 2): S279–88.

Becker, Gary S, and Nigel Tomes. 1976. "Child Endowments and the Quantity and Quality of Children." *Journal of Political Economy* 84 (4, Part 2): S143–62.

Bertrand, Marianne, Emir Kamenica, and Jessica Pan. 2015. "Gender Identity and Relative Income Within Households." *The Quarterly Journal of Economics* 130 (2): 571–614.

Bisin, Alberto, Giorgio Topa, and Thierry Verdier. 2004. "Religious Intermarriage and Socialization in the United States." *Journal of Political Economy* 112 (3): 615–64.

Bisin, Alberto, and Thierry Verdier. 2000. "'Beyond the Melting Pot': Cultural Transmission, Marriage, and the Evolution of Ethnic and Religious Traits." *The Quarterly Journal of Economics* 115 (3): 955–88.

Black, Sandra E, Paul J Devereux, and Kjell G Salvanes. 2005. "The More the Merrier? The Effect of Family Size and Birth Order on Children's Education." *The Quarterly Journal of Economics* 120 (2): 669–700.

Blake, Judith. 1989. *Family Size and Achievement*. Vol. 3. Univ of California Press.

Breen, Richard, and Leire Salazar. 2011. "Educational Assortative Mating and Earnings Inequality in the United States." *American Journal of Sociology* 117 (3): 808–43.

Browning, Martin, Pierre-André Chiappori, and Yoram Weiss. 2014. *Economics of the Family*. Cambridge University Press.

Cabinet Office. 2021. "Basic Data on Marriage and Family (Kekkon to Kazoku Wo Meguru Kiso Data)." https://www.gender.go.jp/kaigi/kento/Marriage-Family/2nd/pdf/1.pdf (Accessed on January 6, 2023).

Cáceres-Delpiano, Julio. 2006. "The Impacts of Family Size on Investment in Child Quality." *Journal of Human Resources* 41 (4): 738–54.

Chernozhukov, Victor, Denis Chetverikov, Mert Demirer, Esther Duflo, Christian Hansen, Whitney Newey, and James M. Robins. 2017. "Double/Debiased Machine Learning for Treatment and Structural Parameters." *Econometrics: Econometric & Statistical Methods - Special Topics eJournal*.

Chiappori, Pierre-Andre, Monica Costa Dias, and Costas Meghir. 2021. "The Measuring of Assortativeness in Marriage: A Comment."

Chiappori, Pierre-André, Sonia Oreffice, and Climent Quintana-Domeque. 2012. "Fatter Attraction: Anthropometric and Socioeconomic Matching on the Marriage Market." *Journal of Political Economy* 120 (4): 659–95.

———. 2016. "Black–White Marital Matching: Race, Anthropometrics, and Socioeconomics." *Journal of Demographic Economics* 82 (4): 399–421.

———. 2018. "Bidimensional Matching with Heterogeneous Preferences: Education and Smoking in the Marriage Market." *Journal of the European Economic Association* 16 (1): 161–98.

Choo, Eugene. 2015. "Dynamic Marriage Matching: An Empirical Framework." *Econometrica* 83 (4): 1373–423.

Cigno, Alessandro, Alessandro Gioffré, and Annalisa Luporini. 2021. "Evolution of Individual Preferences and Persistence of Family Rules." *Review of Economics of the Household* 19 (4): 935–58.

Cigno, Alessandro, Mizuki Komura, and Annalisa Luporini. 2017. "Self-Enforcing Family Rules, Marriage and the (Non) Neutrality of Public Intervention." *Journal of Population Economics* 30 (3): 805–34.

Conley, Dalton, and Rebecca Glauber. 2006. "Parental Educational Investment and Children's Academic Risk Estimates of the Impact of Sibship Size and Birth Order from Exogenous Variation in Fertility." *Journal of Human Resources* 41 (4): 722–37.

Cordón, Juan Antonio Fernández. 1997. "Youth Residential Independence and Autonomy: A Comparative Study." *Journal of Family Issues* 18 (6): 576–607.

Coward, Raymond T, and Jeffrey W Dwyer. 1990. "The Association of Gender, Sibling Network Composition, and Patterns of Parent Care by Adult Children." *Research on Aging* 12 (2): 158–81.

Dwyer, Jeffrey W, and Raymond T Coward. 1991. "A Multivariate Comparison of the Involvement of Adult Sons Versus Daughters in the Care of Impaired Parents." *Journal of Gerontology* 46 (5): S259–69.

Eika, Lasse, Magne Mogstad, and Basit Zafar. 2019. "Educational Assortative Mating and Household Income Inequality." *Journal of Political Economy* 127 (6): 2795–835.

Esping-Andersen, Gøsta. 1997. "Hybrid or Unique?: The Japanese Welfare State Between Europe and America." *Journal of European Social Policy* 7 (3): 179–89.

Eurostat. 2022. "Household Composition Statistics." https://ec.europa.eu/eurostat/statistics-explained (Accessed on January 6, 2023).

Feinberg, Mark E, Anna R Solmeyer, and Susan M McHale. 2012. "The Third Rail of Family Systems: Sibling Relationships, Mental and Behavioral Health, and Preventive Intervention in Childhood and Adolescence." *Clinical Child and Family Psychology Review* 15 (1): 43–57.

Fenton, Norman. 1928. "The Only Child." *The Pedagogical Seminary and Journal of Genetic Psychology* 35 (4): 546–56.

Fernández, Raquel, Alessandra Fogli, and Claudia Olivetti. 2004. "Mothers and Sons: Preference Formation and Female Labor Force Dynamics." *The Quarterly Journal of Economics* 119 (4): 1249–99.

Fernández, Raquel, and Richard Rogerson. 2001. "Sorting and Long-Run Inequality." *The Quarterly Journal of Economics* 116 (4): 1305–41.

Fisman, Raymond, Sheena S Iyengar, Emir Kamenica, and Itamar Simonson. 2006. "Gender Differences in Mate Selection: Evidence from a Speed Dating Experiment." *The Quarterly Journal of Economics* 121 (2): 673–97.

Galor, Oded, and David N Weil. 2000. "Population, Technology, and Growth: From Malthusian Stagnation to the Demographic Transition and Beyond." *American Economic Review* 90 (4): 806–28.

Giuliano, Paola. 2007. "Living Arrangements in Western Europe: Does Cultural Origin Matter?" *Journal of the European Economic Association* 5 (5): 927–52.

Greenwood, Jeremy, Nezih Guner, Georgi Kocharkov, and Cezar Santos. 2014. "Marry Your Like: Assortative Mating and Income Inequality." *American Economic Review* 104 (5): 348–53.

———. 2016. "Technology and the Changing Family: A Unified Model of Marriage, Divorce, Educational Attainment, and Married Female Labor-Force Participation." *American Economic Journal: Macroeconomics* 8 (1): 1–41.

Griffiths, Neil L, Kevin Thomas, Bryce Dyer, Jessica Rea, and Anat Bardi. 2021. "The Values of Only-Children: Power and Benevolence in the Spotlight." *Journal of Research in Personality* 92: 104096.

Hazan, Moshe, and Binyamin Berdugo. 2002. "Child Labour, Fertility, and Economic Growth." *The Economic Journal* 112 (482): 810–28.

Hitsch, Gunter J, Ali Hortaçsu, and Dan Ariely. 2010. "Matching and Sorting in Online Dating." *American Economic Review* 100 (1): 130–63.

Jepsen, Lisa K, and Christopher A Jepsen. 2002. "An Empirical Analysis of the Matching Patterns of Same-Sex and Opposite-Sex Couples." *Demography* 39 (3): 435–53.

Kessler, Daniel. 1991. "Birth Order, Family Size, and Achievement: Family Structure and Wage Determination." *Journal of Labor Economics* 9 (4): 413–26.

Kojima, Hiroshi. 1993. "Sibling Configuration and Coresidence of Married Couples with an Older Mother in Japan." *International Journal of Japanese Sociology* 2 (1): 1–16.

Konrad, Kai A, Harald Künemund, Kjell Erik Lommerud, and Julio R Robledo. 2002a. "Geography of the Family." *American Economic Review* 92 (4): 981–98.

———. 2002b. "Geography of the Family." *American Economic Review* 92 (4): 981–98.

Lee, Jungmin. 2008. "Sibling Size and Investment in Children's Education: An Asian Instrument." *Journal of Population Economics* 21 (4): 855–75.

Li, Bingjing, and Hongliang Zhang. 2017. "Does Population Control Lead to Better Child Quality? Evidence from China's One-Child Policy Enforcement." *Journal of Comparative Economics* 45 (2): 246–60.

Li, Hongbin, Junsen Zhang, and Yi Zhu. 2008. "The Quantity-Quality Trade-Off of Children in a Developing Country: Identification Using Chinese Twins." *Demography* 45 (1): 223–43.

Liu, Haoming. 2014. "The Quality–Quantity Trade-Off: Evidence from the Relaxation of China's One-Child Policy." *Journal of Population Economics* 27 (2): 565–602.

Low, Corinne. 2014. "Essays in Gender Economics." PhD thesis, Columbia University.

Lu, Frances R, and Tom Vogl. 2022. "Intergenerational Persistence in Child Mortality." National Bureau of Economic Research.

Mancillas, Adriean. 2006. "Challenging the Stereotypes about Only Children: A Review of the Literature and Implications for Practice." *Journal of Counseling & Development* 84 (3): 268–75.

Mare, Robert D. 1991. "Five Decades of Educational Assortative Mating." *American Sociological Review*, 15–32.

Mellor, Steven. 1990. "How Do Only Children Differ from Other Children?" *The Journal of Genetic Psychology* 151 (2): 221–30.

Millimet, Daniel L, and Le Wang. 2011. "Is the Quantity-Quality Trade-Off a Trade-Off for All, None, or Some?" *Economic Development and Cultural Change* 60 (1): 155–95.

Ministry of Health, Labour and Welfare. 2020. "Summary Report of Comprehensive Survey of Living Conditions 2019." https://www.mhlw.go.jp/english/database/db-hss/dl/report_gaikyo_2019.pdf (Accessed on January 6, 2023).

Moav, Omer. 2005. "Cheap Children and the Persistence of Poverty." *The Economic Journal* 115 (500): 88–110.

Ogawa, Naohiro, and John F Ermisch. 1996. "Family Structure, Home Time Demands, and the Employment Patterns of Japanese Married Women." *Journal of Labor Economics* 14 (4): 677–702.

Ogawa, Naohiro, and Robert D Retherford. 1993. "Care of the Elderly in Japan: Changing Norms and Expectations." *Journal of Marriage and the Family*, 585–97.

Pencavel, John. 1998. "Assortative Mating by Schooling and the Work Behavior of Wives and Husbands." *The American Economic Review* 88 (2): 326–29.

Polit, Denise F, and Toni Falbo. 1987. "Only Children and Personality Development: A Quantitative Review." *Journal of Marriage and the Family*, 309–25.

———. 1988. "The Intellectual Achievement of Only Children." *Journal of Biosocial Science* 20 (3): 275–86.

Polit, Denise F, Ronald L Nuttall, and Ena V Nuttall. 1980. "The Only Child Grows up: A Look at Some Characteristics of Adult Only Children." *Family Relations*, 99–106.

Poston Jr, Dudley L, and Toni Falbo. 1990. "Academic Performance and Personality Traits of Chinese Children:" Onlies" Versus Others." *American Journal of Sociology* 96 (2): 433–51.

Qian, Nancy. 2009. "Quantity-Quality and the One Child Policy: The Only-Child Disadvantage in School Enrollment in Rural China." National Bureau of Economic Research.

Qin, Xuezheng, Castiel Chen Zhuang, and Rudai Yang. 2017. "Does the One-Child Policy Improve Children's Human Capital in Urban China? A Regression Discontinuity Design." *Journal of Comparative Economics* 45 (2): 287–303.

Rainer, Helmut, and Thomas Siedler. 2009. "O Brother, Where Art Thou? The Effects of Having a Sibling on Geographic Mobility and Labour Market Outcomes." *Economica* 76 (303): 528–56.

———. 2012. "Family Location and Caregiving Patterns from an International Perspective." *Population and Development Review* 38 (2): 337–51.

Raymo, James M. 2003. "Educational Attainment and the Transition to First Marriage Among Japanese Women." *Demography* 40 (1): 83–103.

Raymo, James M, and Hiromi Ono. 2007. "Coresidence with Parents, Women's Economic Resources, and the Transition to Marriage in Japan." *Journal of Family Issues* 28 (5): 653–81.

Reese-Weber, Marla, and Jeffrey H Kahn. 2005. "Familial Predictors of Sibling and Romantic-Partner Conflict Resolution: Comparing Late Adolescents from Intact and Divorced Families." *Journal of Adolescence* 28 (4): 479–93.

Robins, James M, and Andrea Rotnitzky. 1995. "Semiparametric Efficiency in Multivariate Regression Models with Missing Data." *Journal of the American Statistical Association* 90 (429): 122–29.

Rosenzweig, Mark R, and Kenneth I Wolpin. 1980. "Testing the Quantity-Quality Fertility Model: The Use of Twins as a Natural Experiment." *Econometrica: Journal of the Econometric Society*, 227–40.

Rosenzweig, Mark R, and Junsen Zhang. 2009. "Do Population Control Policies Induce More Human Capital Investment? Twins, Birth Weight and China's 'One-Child' Policy." *The Review of Economic Studies* 76 (3): 1149–74.

Sakamoto, Kazuyasu, and Yukinobu Kitamura. 2007. "Marriage Behavior from the Perspective of Intergenerational Relationships." *Japanese Economy* 34 (4): 76–122.

Semenova, Vira, and Victor Chernozhukov. 2021. "Debiased Machine Learning of Conditional Average Treatment Effects and Other Causal Functions." *The Econometrics Journal* 24 (2): 264–89.

Siow, Aloysius. 2015. "Testing Becker's Theory of Positive Assortative Matching." *Journal of Labor Economics* 33 (2): 409–41.

Spitze, Glenna, and John R Logan. 1991. "Sibling Structure and Intergenerational Relations." *Journal of Marriage and the Family*, 871–84.

Suitor, J Jill, and Karl Pillemer. 2007. "Mothers' Favoritism in Later Life: The Role of Children's Birth Order." *Research on Aging* 29 (1): 32–55.

The Dai-ichi Life Research Institute. 2007. "The Summary Report on the Results of Survey on Inheritance Among Middle-Aged and Older Adults in Japan, 2005 (Japanese)." https://www.dlri.co.jp/pdf/ld/01-14/news0701.pdf.

The Office for National Statistics. 2020. "Number of Families by Number of Dependent Children, UK, 1996 to 2019." https://www.ons.gov.uk/peoplepopulationandcommunity/birthsdeathsandmarriages/families (Accessed on January 6, 2023).

The Yu-cho Foundation. 2023. "The Summary Report on the Results of the 5th Survey on Household Finances and Savings in Japan, 2022 (Japanese)." https://www.yu-cho-f.jp/wp-content/uploads/survey_report-9.pdf.

Uchikoshi, Fumiya, James M Raymo, and Shohei Yoda. 2023. "Family Norms and Declining First-Marriage Rates: The Role of Sibship Position in the Japanese Marriage Market." *Demography*, 10741873.

Vogl, Tom S. 2013. "Marriage Institutions and Sibling Competition: Evidence from South Asia." *The Quarterly Journal of Economics* 128 (3): 1017–72.

———. 2020. "Intergenerational Associations and the Fertility Transition." *Journal of the European Economic Association* 18 (6): 2972–3005.

Wakabayashi, Midori, and Charles Yuji Horioka. 2009. "Is the Eldest Son Different? The Residential Choice of Siblings in Japan." *Japan and the World Economy* 21 (4): 337–48.

Whiteman, Shawn D, Susan M McHale, and Ann C Crouter. 2003. "What Parents Learn from Experience: The First Child as a First Draft?" *Journal of Marriage and Family* 65 (3): 608–21.

Wu, Jiabin, and Hanzhe Zhang. 2021. "Preference Evolution in Different Matching Markets." *European Economic Review* 137: 103804.

Yu, Wei-hsin, and Ekaterina Hertog. 2018. "Family Characteristics and Mate Selection: Evidence from Computer-Assisted Dating in Japan." *Journal of Marriage and Family* 80 (3): 589–606.

Yu, Wei-hsin, and Janet Chen-Lan Kuo. 2016. "Explaining the Effect of Parent-Child Coresidence on Marriage Formation: The Case of Japan." *Demography* 53 (5): 1283–1318.

Yu, Wei-hsin, Kuo-hsien Su, and Chi-Tsun Chiu. 2012. "Sibship Characteristics and Transition to First Marriage in Taiwan: Explaining Gender Asymmetries." *Population Research and Policy Review* 31 (4): 609–36.